\RequirePackage{amsthm} 

\documentclass[sn-basic,iicol,pdflatex]{sn-jnl}

\usepackage[utf8]{inputenc}
\usepackage[T1]{fontenc}
\usepackage{float}
\usepackage{amsmath,amssymb,amsfonts}
\usepackage{mathrsfs}
\usepackage{theorem}
\usepackage{graphicx}
\usepackage{color}
\usepackage{wasysym}
\usepackage{hyperref} 
\usepackage{natbib}

\usepackage{mathtools}
\usepackage{dcolumn}
\usepackage{multirow}

\usepackage[title]{appendix}%
\usepackage{xcolor}%
\usepackage{textcomp}%
\usepackage{manyfoot}%
\usepackage{booktabs}%
\usepackage[switch]{lineno} 
\usepackage{datetime2}

\usepackage[finalnew]{trackchanges} 

\begin{document}

%

\newcommand{\ama}[1]{{\color{ama} #1}}
\newcommand{\vio}[1]{{\color{vio} #1}}
\newcommand{\blue}[1]{{\color{blue} #1}}
\newcommand{\green}[1]{{\color{green} #1}}
\newcommand{\red}[1]{{\color{red} #1}}
\newcommand{\marron}[1]{{\color{marron} #1}}
\newcommand{\verde}[1]{{\color{verde} #1}}
\definecolor{verde}{rgb}{0.,.5,0.4}
\definecolor{blue}{rgb}{0,0,1}
\definecolor{green}{rgb}{0,0.65,0.5}
\definecolor{marron}{rgb}{0.7,0.2,0.1}
\definecolor{red}{rgb}{1,0,0}
\definecolor{vio}{rgb}{0.66,0,1}
\definecolor{ama}{rgb}{1,1,0}



\title[Localization of GW190521 ...]{\sf 
	Localization of GW190521 and reconstruction of 
	the spin-2 gravitational-wave polarization modes
}

\author[1,2]{\fnm{Osvaldo M.} \sur{Moreschi}
	\href{https://orcid.org/0000-0001-9753-3820}{\includegraphics[scale=0.4]{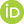}} }\email{o.moreschi@unc.edu.ar}

%

\affil[1]{\orgdiv{Facultad de Matemática, Astronomía, Física y Computación(FaMAF)}, 
	\orgname{Universidad Nacional de Córdoba}, 
	\orgaddress{\street{Ciudad Universitaria}, \city{Córdoba}, \postcode{(5000)}, 
		\country{Argentina}}}

\affil[2]{\orgdiv{Instituto de Física Enrique Gaviola(IFEG)}, 
	\orgname{CONICET}, 
	\orgaddress{\street{Ciudad Universitaria}, \city{Córdoba}, \postcode{(5000)}, 
		\country{Argentina}}}


\abstract{
	
We report on the new developments of the procedure L2D+PMR 
for sky localization\citep{KAGRA:2013rdx} using data from 
two gravitational-wave detectors and the reconstruction of 
gravitational-wave polarization modes\citep{Poisson2014}, which we have previously 
presented\citep{Moreschi25a}. 
In this case we apply our methods to the GW190521\citep{Abbott:2020tfl} 
event, which was recorded by three gravitational-wave observatories. 
We find the localization of the source close to one of the crossings
of two delay rings, as expected.
Thus, our results corroborate the consistency of the procedure.
We compare 
our findings with those obtained by the LIGO/Virgo Collaborations, 
providing an independent assessment of gravitational-wave localization. 
We present the second direct measurement of spin-2 gravitational-wave 
polarization modes.

}

%


\maketitle

\date{\DTMnow} 


{\sf \footnotesize 
\tableofcontents
}


\section{Introduction}

This is the second article in a series dedicated to the application and examination of the 
L2D+PMR procedure, originally introduced in \cite{Moreschi25a}.
This method enables the sky localization of a gravitational-wave source using data 
from only two observatories while also allowing for the reconstruction of the spin-2 
polarization modes (PMs) of the detected gravitational wave (GW).
By further validating and refining this procedure, we aim to demonstrate 
its robustness and applicability to real gravitational-wave events.

The problem of localizing the sky position of a gravitational-wave source \citep{KAGRA:2013rdx} 
changes significantly when more than two detectors record the signal, 
compared to the case where only two detectors are involved.
This is the case of the event GW190521\citep{Abbott:2020tfl}, where the two LIGO, Hanford and Livingston
and the Virgo detectors have recorded the signals.
This allows for improving the basic localization technique supported on
the information of time delays among the observatories.
With only two detectors, the source is typically constrained to a ring-like region on the celestial sphere, 
defined by the possible arrival-time delays between the two observatories. 
However, with a third detector, the localization is improved: instead of an extended ring, 
the source is now constrained to two possible points, determined by the intersection of two delay rings. 
This improved localization provides a stronger foundation for multi-messenger follow-up 
observations and enhances the astrophysical interpretation of the event.

Therefore, if an independent localization method is applied, it should yield a position 
close to one of the two possible points determined by the delay ring intersections. 
This is precisely one of the key results we present in this article.
When applying the L2D+PMR procedure, which we previously introduced in \citep{Moreschi25a}, 
to the event GW190521, our method provides a localization close to the southern intersection 
point of the Hanford–Livingston (H-L) delay ring and the Virgo–Livingston (V-L) delay ring. 
This result is in excellent agreement with expectations and serves as a strong corroboration 
of the accuracy and reliability of our method.

Our techniques do not rely on the parameters of a specific physical model of the source. 
Instead, they are based purely on the fundamental spin-2 polarization properties 
of GWs\citep{Eardley:1973br, Eardley:1974nw, Poisson2014}. 
This intrinsic approach allows for a more model-independent analysis, avoiding biases 
introduced by assumed astrophysical parameters.
As a result, our method not only enables precise source localization using just two detectors 
but also facilitates the direct reconstruction of the two PMs of 
the gravitational-wave signal. 
This is a crucial result, as it allows for a direct observational test of
General Relativity’s prediction that GWs of astrophysical origin exhibit only 
two tensorial polarization states(See discussion in \cite{Moreschi25a}.).
Accordingly, in this work, we present the second-ever direct measurement of 
the + (plus) and $\times$  (cross) polarization modes of a GW.

This article focuses on the application of the L2D+PMR procedure to the GW190521 event. 
Rather than explore the detailed astrophysical parameters of a specific source model, 
our primary objective is to provide localization results and PM
reconstruction using only the fundamental theoretical framework of a spin-2 GW.
This approach ensures that our results remain as model-independent as possible.

The GW190521 event, which was observed during the O3a LIGO/Virgo run,
has strong enough signals which allow us to apply
our procedure, and enables us to not only perform an independent localization 
of the source but also to extract the spin-2 PM
of the detected GW.

In \cite{LIGOScientific:2020ufj} the LIGO and Virgo Collaborations
presented a detailed analysis of the gravitational-wave signals of the GW190521 event,
and concluded that it was consistent with a binary black hole merger at
redshift 0.8 with high component masses of $85M_\odot$ and $66M_\odot$.
However these values were later updated on May 13, 2022, 
on the official event webpage
\href{https://gwosc.org/eventapi/html/GWTC-2.1-confident/GW190521/v4/}{gwosc.GW190521},
which revised the estimates to a redshift of 0.56 with component masses of $98.4M_\odot$ and $57.2M_\odot$.

In this work, we compare our localization results with those reported in \cite{LIGOScientific:2020ufj}, 
providing an independent assessment of the sky position of the source. 
The subject of localization has also been explored in \cite{Szczepanczyk:2020osv}, 
where their sky maps exhibit localization regions that closely resemble those reported by 
the LIGO/Virgo Collaborations.

This article is organized as follows.
In section \ref{sec:GWintermsofPM} we recall the basic equation that relates the recorded
signal in each gravitational-wave detector with the spin-2 polarization modes.
The characteristics of the strain for this event are mentioned in section \ref{sec:charact-strain}.
The way in which we obtained the relatives time delays for this event
is described in section \ref{sec:time-delays}.
In section \ref{sec:denoising} we present the results of applying the denoising techniques
to this case.
The study of the strains in the time-frequency domain is presented in
section \ref{sec:time_freq}; where both scalograms are shown.
The description of the universal fitting process along with its results
is presented in section \ref{sec:univ_fitt_chirp}.
The localization of the source of GW190521 is shown in section \ref{sec:loc}
along with its comparison with LIGO results.
The reconstruction of the gravitational-wave PM for this event
is presented in section \ref{sec:PM}; which is shown in several
polarization frames.
In section \ref{sec:rec-signalfromPM} we test the possible content in this strains
from other spin PMs.
We reserve section \ref{sec:final} for final comments.

\section{Gravitational wave in terms of the spin-2 polarization modes}\label{sec:GWintermsofPM}

The content of a gravitational-wave signal detected by an observatory $X$ is shaped 
by the detector’s response, which is modulated by its antenna pattern functions. 
These functions account for the detector's sensitivity to different polarization 
modes depending on the source's sky position, inclination, and polarization angle. 
This modulation is mathematically described in Equation \ref{eq:vX}, where the 
recorded signal is expressed as a linear combination of the two spin-2 polarization 
modes weighted by the detector's pattern functions.
The pattern functions depend on the relative orientation of the detector's arms 
with respect to the incoming GW, which introduces directional 
dependence into the observed signal; and also depend on the angle of
the polarization frame, discussed in \cite{Moreschi25a}. 
As a result, different detectors 
will record a particular projection of the waveform amplitudes, allowing for source localization 
and the potential reconstruction of polarization modes.
In this way, the strain $v_X$ recorded at detector $X$ can be expressed by
\begin{equation}\label{eq:vX}
\begin{split}
v_X(t + \tau_X) &= n_X(t + \tau_X) + s_X(t + \tau_X) \\
&= n_X(t + \tau_X) + 
F_{+X}(\theta_X,\phi_X,\psi_X,t) s_+(t) \\
&\;\;\; +
F_{\times X}(\theta_X,\phi_X,\psi_X,t) s_\times(t)
,
\end{split}
\end{equation}
where $X$ stands for $H$(Hanford) or $L$(Livingston), $\tau_X$ is the delay of detector $X$ with respect to
the chosen reference time, $(\theta_X,\phi_X)$ are the angular coordinates with respect
to detector $X$ of the direction of the source, $\psi_X$ is the angle of the
GW frame and $t$ is the time.
The strain is denoted by $v$, we use $n$ to refer to the noise, $s$ for the signal,
which is decomposed in the PMs $s_+$ and $s_\times$;
while $F_+$ and $F_\times$ are the detector pattern functions\citep{Moreschi25a}.

It can be seen from eq. \ref{eq:vX}, that for any meaningful discussion of the PMs,
it is essential to have estimates of the signal delay times, between the observatories.
These delays, arising due to the finite speed of the signals, provide essential timing information 
that is used to locate the source in the sky.
This issue will be addressed in Section \ref{sec:time-delays}, where we will discuss 
the methods for determining these delays and their importance for improving 
the localization precision of gravitational-wave sources. 
Properly accounting for time delays allows for a more accurate interpretation 
of the detected signal, especially when multiple detectors are involved.

\section{Characteristics of the strain}\label{sec:charact-strain}

GW190521 is classified as a GWTC-2.1-confident event on the official GWOSC 
page \href{https://gwosc.org/eventapi/html/GWTC-2.1-confident/GW190521/v4/}{gwosc.org}, 
but for our analysis, we use the v1 strain data presented in the O3\_Discovery\_Papers. 
The GPS reference time for the event is 1242442967.5, which differs by 0.1s from 
the LIGO reference time to include some additional physical signal beyond their chosen reference. 
The corresponding UTC time is 2019-05-21 03:02:29.5.

This event was assigned a network SNR (signal-to-noise ratio) of 14.3 and  
a sky localization area of 1000 square degrees. 
The redshift of the event is 0.56, reflecting the updated value as of May 13, 2022. 
The signal was detected by the two LIGO detectors (Hanford and Livingston) as well as Virgo.

For the analysis, we use 16384Hz sampled data, which provides the necessary time resolution 
for our procedure.

In \cite{Abbott:2020tfl}, the GW190521 event was presented as a short transient signal 
with a duration of approximately 0.1s. However, for our analysis, we used a time window 
of 0.35s for the LIGO detectors and 0.18s for the Virgo detector. 
This longer window was chosen to better capture the full dynamics of the event.
The authors of \cite{Abbott:2020tfl} also mentioned the appearance of around four cycles 
in the frequency range of 30–80 Hz during the event. 
Despite this, no clear signal is visible in Figure 1 of their publication, 
both in the time domain and time-frequency domain graphs, for the Virgo strain data.


We apply the filtering techniques described in \cite{Moreschi:2019vxw} to 
the three strains from the LIGO and Virgo detectors. 
First, we use a general bandpass filter with frequencies ranging from 25.0Hz to 995.0Hz.
Following this, we apply specific stopband filters tailored to each detector 
to further mitigate noise in each detector, 
ensuring that the gravitational-wave signal remains the dominant feature in the frequency band of interest.
These filtering steps are essential for ensuring that the data from each detector are in optimal form 
for subsequent analysis, such as source localization and polarization mode reconstruction. 

\section{Time delays for GW190521}\label{sec:time-delays}

\begin{figure}[H]
\centering
\includegraphics[clip,width=0.5\textwidth]{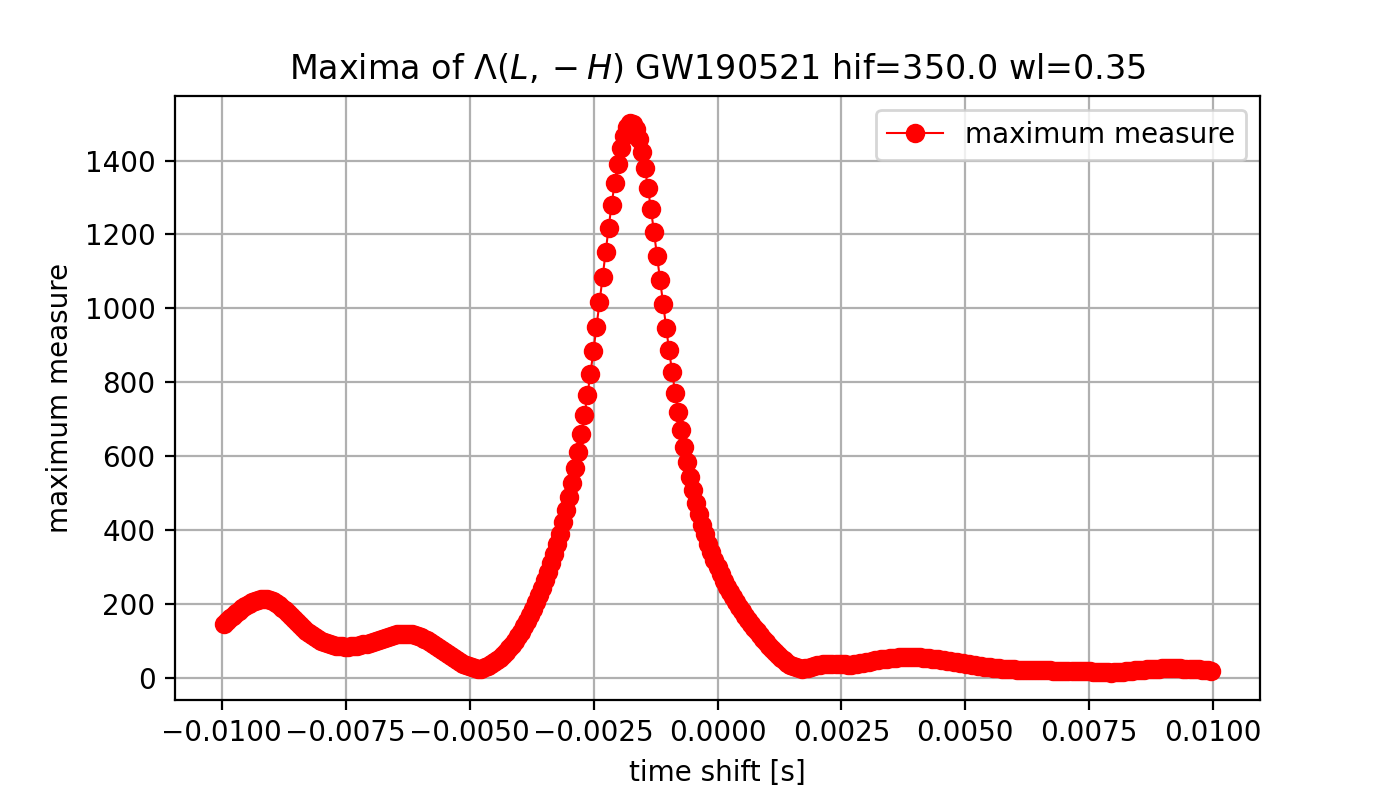}
\caption{This graph shows the behavior of mathematical evaluation $\Lambda$ of
	the OM measure as a function of shift
	of the Hanford data with respect to the Livingston data for a window of 0.35s and
	a limiting high frequency filter at 350Hz.
}
\label{fig:lieklihoodH}
\end{figure}

The relative delay times among the observatories for the GW190521 event
have not been publicly released, 
and determining them proved to be a challenging task.
Our procedure to determine the time delay of the strain of Hanford (H) with respect to Livingston (L)
is to use the optimized measure OM\citep{Moreschi:2024njx} with an appropriate window of length $wl$.
In a preliminary study of the signals we choose initially $wl=0.35$s; but the signal was
very noisy. Then we repeated the study with half of this window length and found
a local maximum at $t_{dH0}=-0.001404$s. Then we studied also the strain with a lowpass
filter at 350Hz which resulted in a slightly different maximum at $t_{dH}=-0.001770$s.
We adopt this latter value as the nominal time delay for H with respect to L.
Figure \ref{fig:lieklihoodH} presents the corresponding 
mathematical evaluation $\Lambda$ 
of the OM measure as a function of the relative time shift, 
illustrating the optimal delay determination. 
This precise estimation of time delays is crucial for sky localization and 
polarization mode reconstruction, as even small uncertainties can significantly affect the results.

\begin{figure}[H]
\centering
\includegraphics[clip,width=0.5\textwidth]{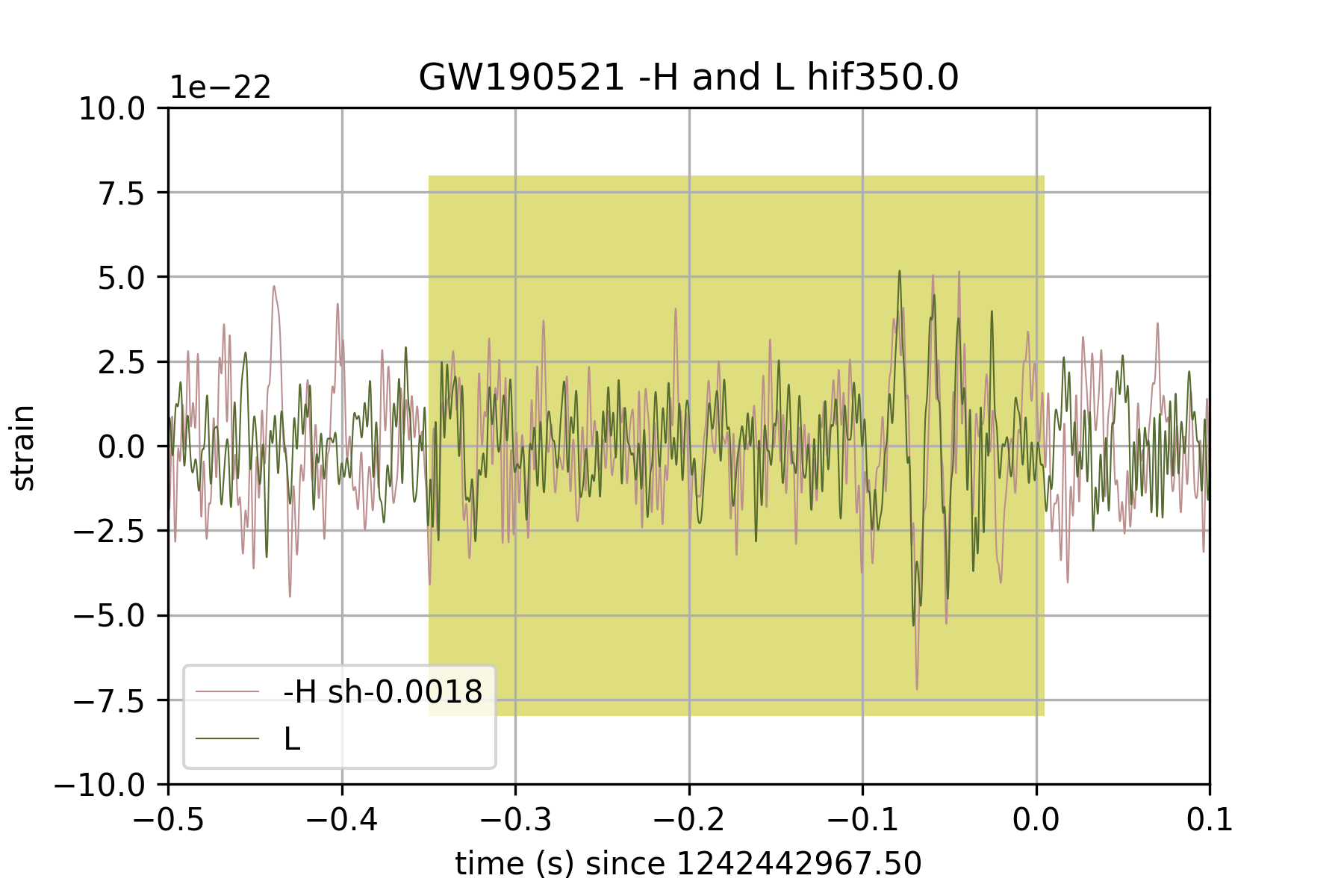}
\caption{Strains of -H and L with a	limiting high frequency filter at 350Hz and with $t_{dH}=-0.001770$s.
	The colored rectangular region indicates the initial width of the window used 
	in the measure. 
}
\label{fig:-HyLat350Hz}
\end{figure}

In order to confirm that the maximum shown in \ref{fig:lieklihoodH} really
corresponds to a coincidence of signals near the nominal time of the event,
we show in Fig. \ref{fig:-HyLat350Hz} the comparison of the shifted -H strain
with the L strain, both filtered with a 350Hz low-pass filter.
This visualization highlights the similarity between the two signals in 
the vicinity of the reference time for GW190521, reinforcing the validity 
of the estimated H–L delay time. 
The application of a low-pass filter helps suppress high-frequency noise, 
making the underlying signal structure more apparent. 
This provides additional support for the accuracy of our delay measurement,
ensuring that it is not an artifact of noise but rather a genuine 
feature of the detected gravitational-wave signal.

\begin{figure}[H]
\centering
\includegraphics[clip,width=0.5\textwidth]{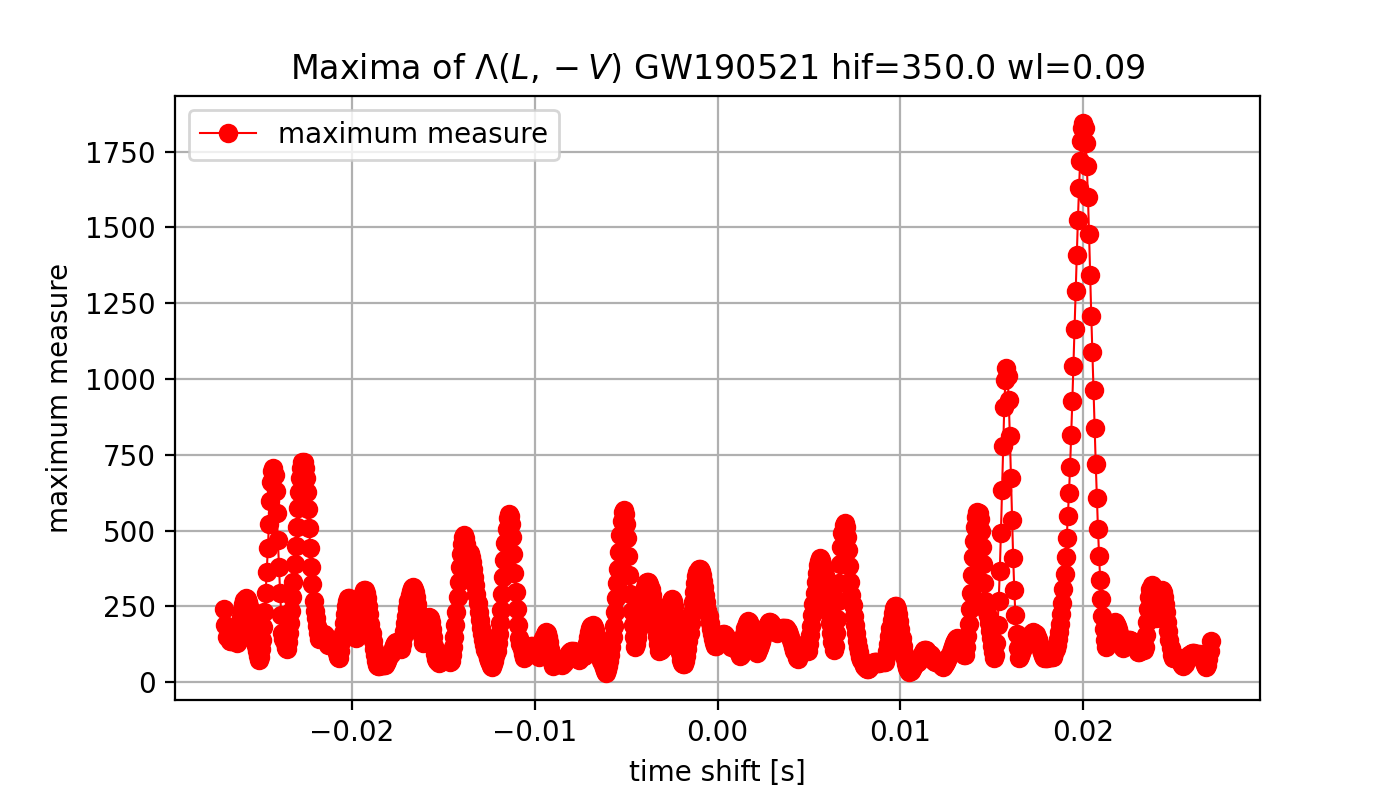}
\caption{This graph shows the behavior of the OM measure as a function of shift
	of the Virgo data with respect to the Livingston data for a window of 0.09s and
	a limiting high frequency filter at 350Hz.
}
\label{fig:lieklihoodV}
\end{figure}

The analysis of the Virgo (V) strain proved to be more challenging. 
We examined the strain using various lowpass filters and tested different 
window lengths for optimal signal extraction.
By applying an 80 Hz low-pass filter and using a window length of $wl=0.18$s, 
we identified a local maximum in the OM measure at $t_{dV0} = 0.01514$s;
which corresponds to a similar behavior for the strains around $-0.2$s of the time
of maximum amplitudes.
We also find the maximum of the OM, for a limiting high frequency filter at 350Hz with half of $wl$,
with a shift of $t_{dV}=0.02002$s.
This shift aligns with signal structures at the time of maximum amplitudes, 
making it the preferred estimate for the V–L delay time.
In Fig. \ref{fig:lieklihoodV} we present the $\Lambda$ measure as a function of the relative time shift, 
between the Virgo data with the Livingston strain, showing the best time delay.
One can see in Fig. \ref{fig:lieklihoodV} that the $\Lambda$ measure 
is much more noisy due to the nature of the Virgo strain.

\begin{figure}[H]
\centering
\includegraphics[clip,width=0.5\textwidth]{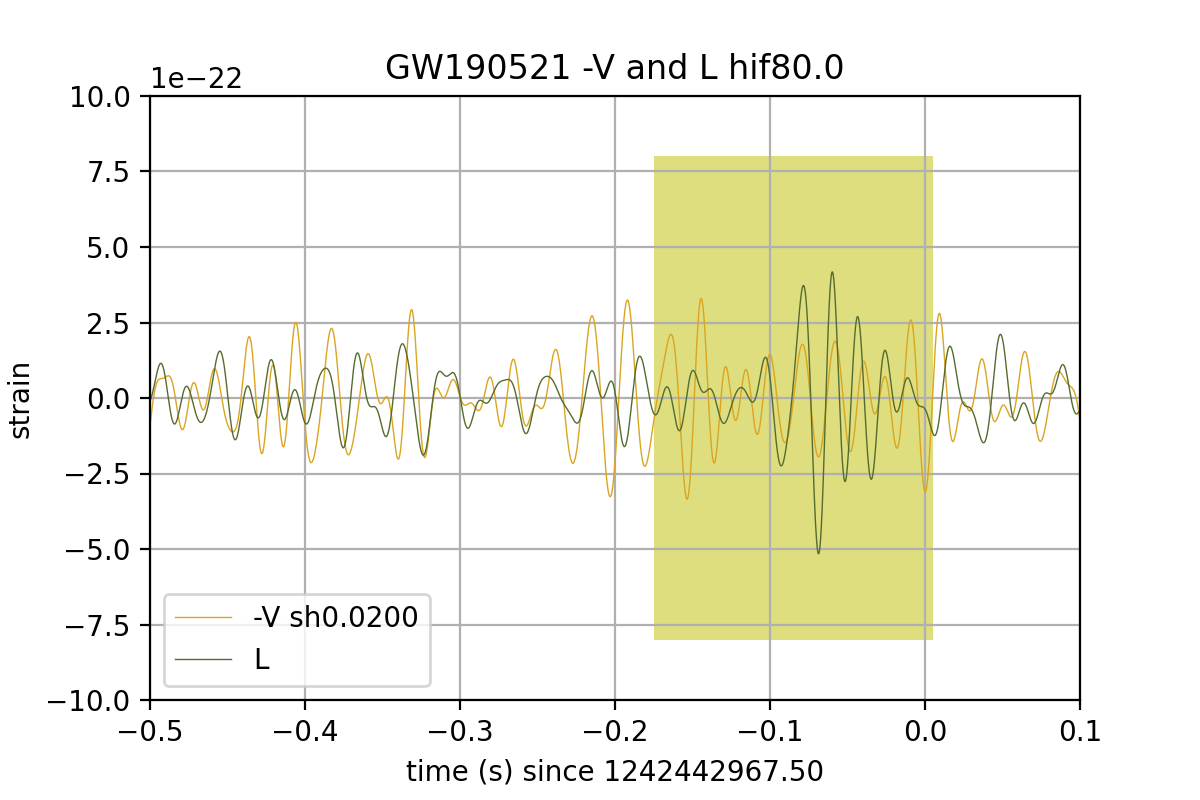}
\caption{Strains of -V and L with a	limiting high frequency filter at 80Hz and with $t_{dV}= 0.02002$s.
	The colored rectangular region indicates the initial width of the window used 
in the measure; although for the $\Lambda$ curve shown in  Fig. \ref{fig:lieklihoodV}
half of this window was used. 
}
\label{fig:-VyLat80Hz}
\end{figure}

To check that this time delay determination actually gives a coincidence of 
the signals in both strains, in Fig. \ref{fig:-VyLat80Hz} we present a comparison of the -V strain with 
the L strain, both filtered at 80 Hz. A noticeable similarity between 
the signals near the GW190521 reference time further supports our delay determination.

Using the operational time delays derived earlier, along with the information of 
the high-frequency 
content of the signal, which in this case we take $\nu_{max} = 100$Hz, we estimate 
a parameter $\sigma$. This parameter allows us to define Gaussian distributions
around the time delay rings, as outlined in our previous companion article.

\begin{figure}[H]
\centering
\includegraphics[clip,width=0.5\textwidth]{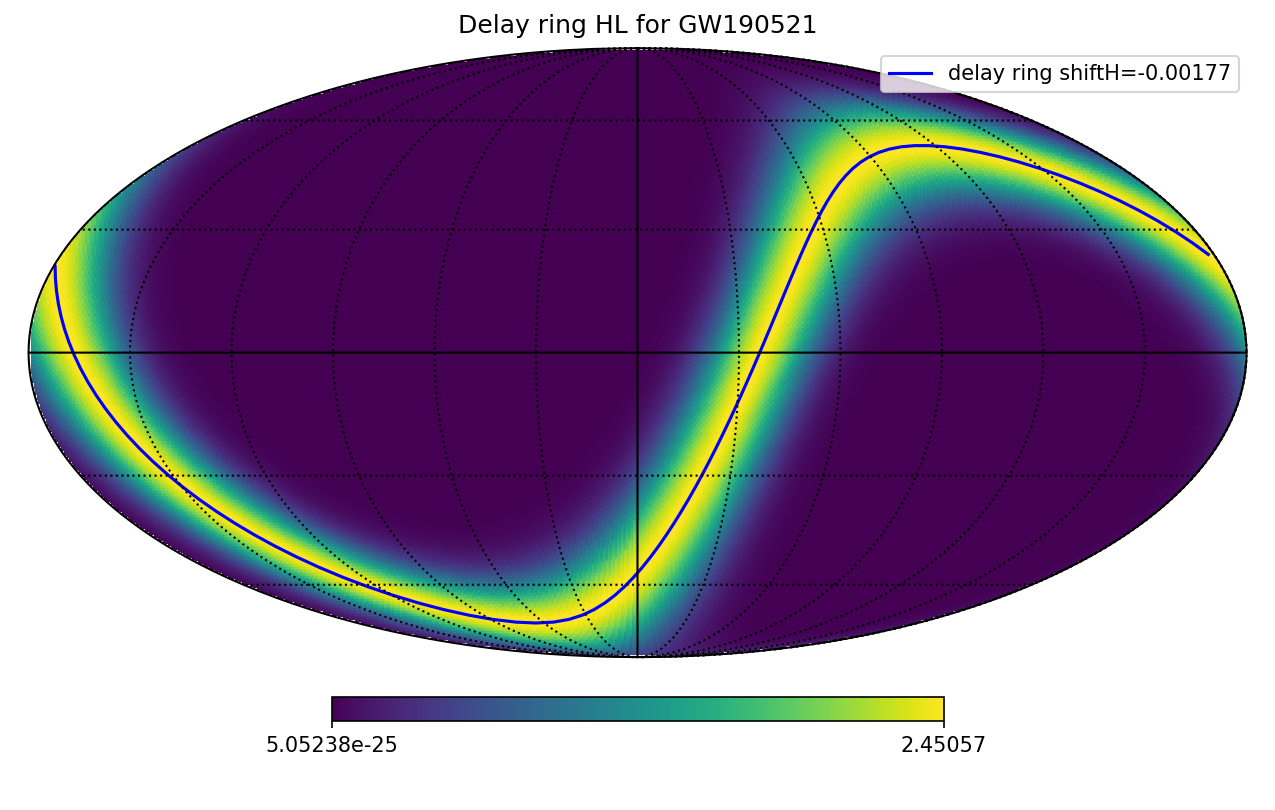}
\caption{Delay ring for Hanford using a Mollweide projection in equatorial coordinate system
	with origin at the center and East towards left.
	It is also shown the corresponding Gaussian region.
}
\label{fig:ringH}
\end{figure}

Figure \ref{fig:ringH} illustrates the nominal time delay ring for Hanford 
relative to Livingston, along with its corresponding Gaussian distribution. 
This visualization highlights the expected sky localization constraints based on 
time delay studies.

The width of the Gaussian region is primarily influenced by the low maximum frequency 
content of the gravitational-wave signal. Lower frequencies result in greater 
timing uncertainties, leading to broader localization regions. 
However, despite this limitation, the area shown in Fig. \ref{fig:ringH} 
serves as a first estimate for the source's position.

\begin{figure}[H]
\centering
\includegraphics[clip,width=0.5\textwidth]{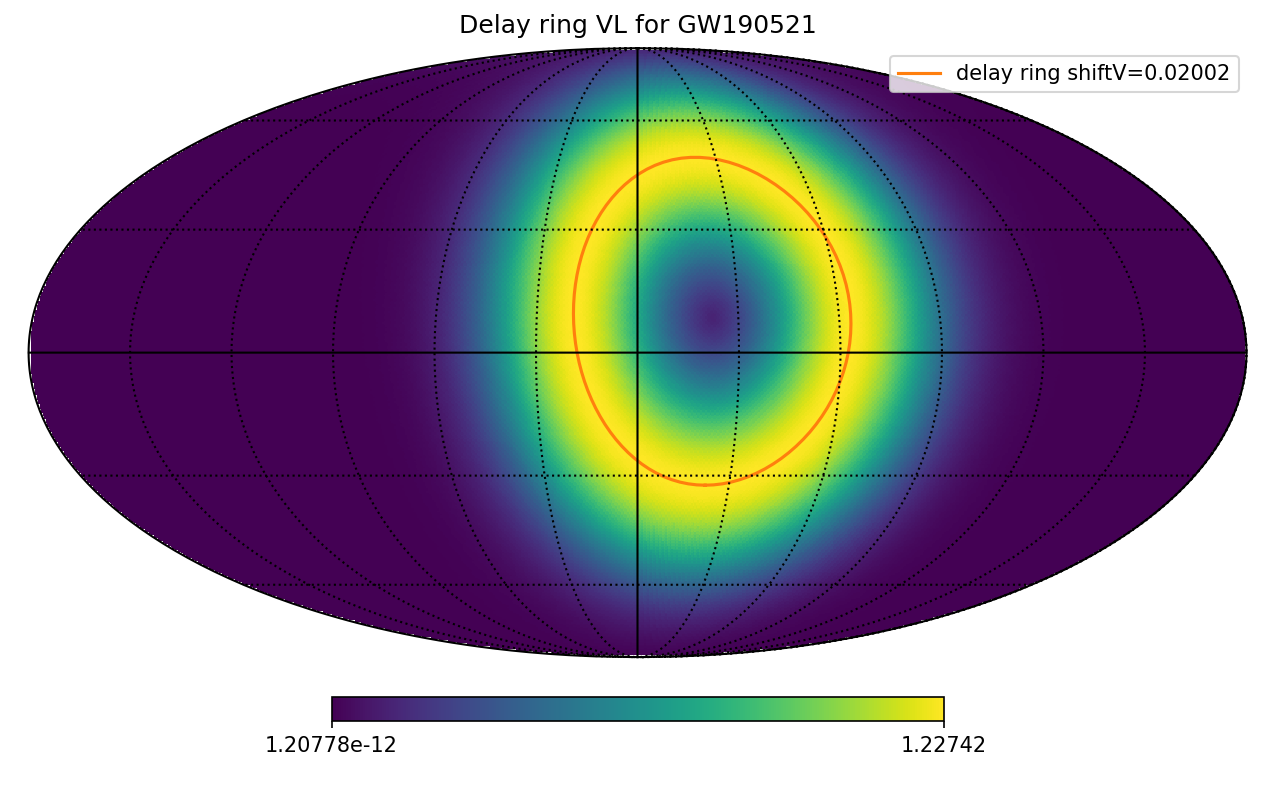}
\caption{Delay ring for Virgo, and corresponding Gaussian region.
}
\label{fig:ringV}
\end{figure}

For the (-V, L) time delay ring, the corresponding localization region 
is displayed in Fig. \ref{fig:ringV}. In this case, the region appears 
wider due to the increased noise levels in the strain V, 
which affects the precision of the estimated time delay and consequently 
the localization accuracy.

Each Gaussian region around the delay rings represents an estimate of 
the probability distribution for the source's location in the sky. 
Since these regions correspond to independent pieces of information, 
the optimal localization estimate is obtained by taking the product of both probability distributions.

\begin{figure}[H]
\centering
\includegraphics[clip,width=0.5\textwidth]{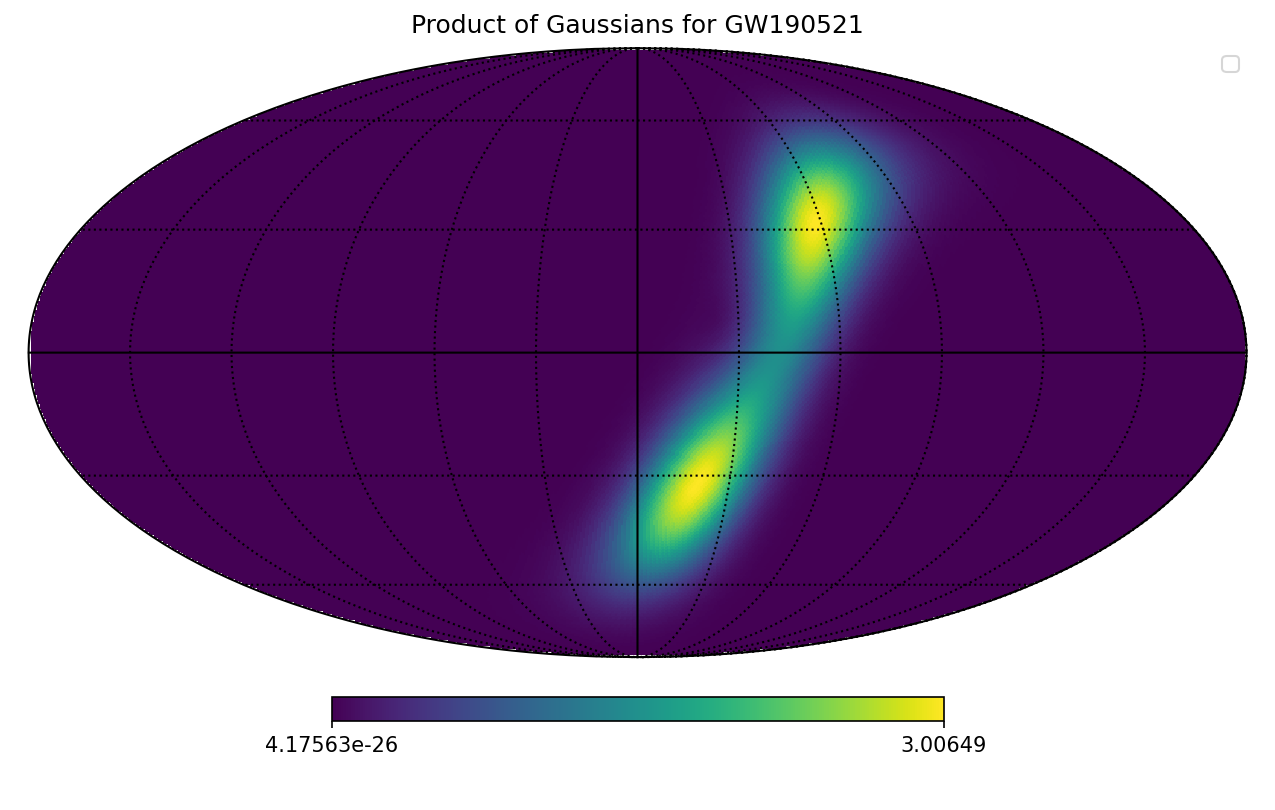}
\caption{Estimate of the probability distribution for the localization of the
	gravitational-wave source, considering the information encoded in 
	the measurement of the two delay rings.
}
\label{fig:prod-gusHxgausV}
\end{figure}

The resulting combined distribution, shown in Fig. \ref{fig:prod-gusHxgausV}, 
provides a more refined triangulation of the source position. 
This triangulation method is commonly used in gravitational-wave astronomy 
to improve sky localization by incorporating information from multiple detectors.
Thus, Fig. \ref{fig:prod-gusHxgausV} represents our best localization 
estimate prior to implementing the full L2D+PMR procedure as described 
in our previous companion article. At this stage, our localization 
is derived solely from time-delay triangulation and the corresponding 
Gaussian probability regions, providing a preliminary but solid 
constraint on the source’s position.

The subsequent application of the L2D+PMR procedure will refine this 
localization further, incorporating additional information to improve 
precision and reliability.

\section{Estimates of the signals by wavelet denoising}\label{sec:denoising}

Applying wavelet denoising techniques, as described in \cite{Moreschi25a}, 
is a crucial step in the procedure. This method allows 
for more accurate estimation of the underlying waveform.
Thus, we obtain the following denoised estimates of the signals, which 
serve as the foundation for the subsequent localization and polarization mode reconstruction steps:
\begin{equation}\label{eq:wX}
\begin{split}
w_X(t + \tau_X) &= e_X(t + \tau_X) + s_X(t + \tau_X) \\
&= e_X(t + \tau_X) + 
F_{+X0}(\theta_X,\phi_X,\psi_X,t) s_+(t)\\
&\;\;\;+
F_{\times X0}(\theta_X,\phi_X,\psi_X,t) s_\times(t)
,
\end{split}
\end{equation}
where $w_X$ are the estimates, and now $e_X$ stands 
for the error intrinsic to the estimates.
Contrary to the previous situation, now we assume that the magnitude of the
errors are much smaller than the magnitude of the signals.
We will also assume that the scalar product of the error with the signals
are negligible.

\begin{figure}[H]
\centering
\includegraphics[clip,width=0.5\textwidth]{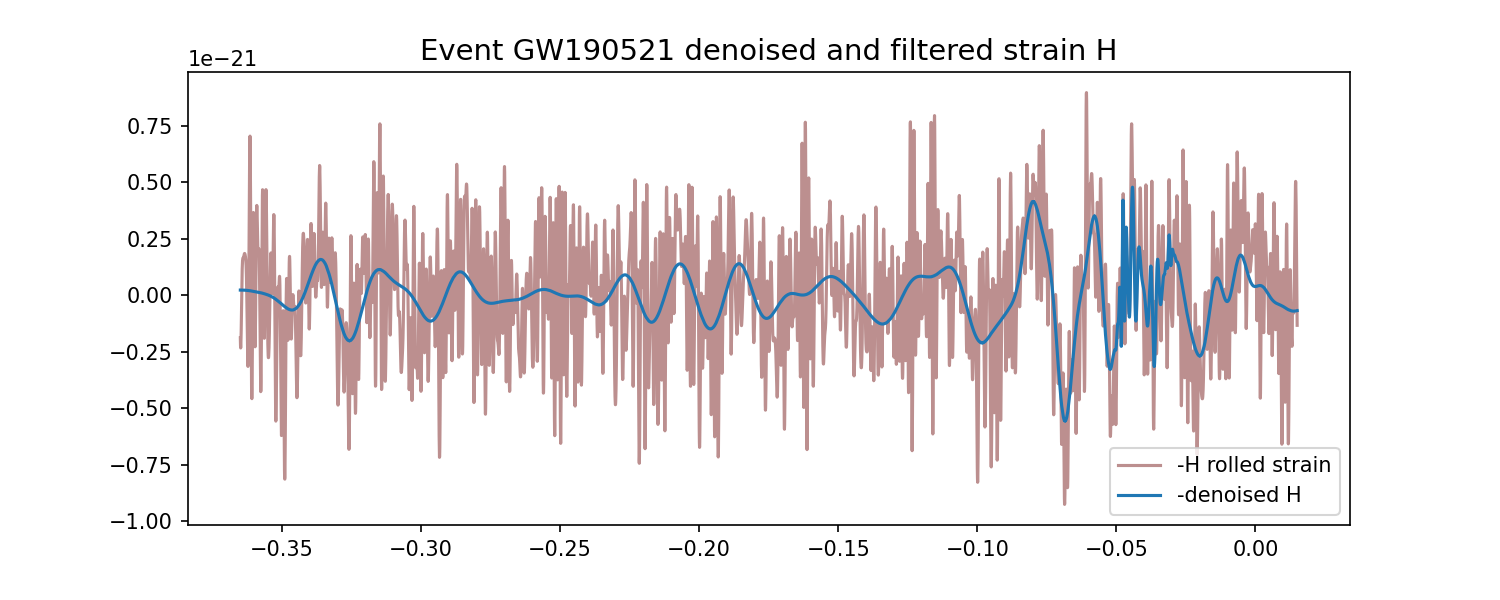}
\caption{Strain and denoised data for Hanford LIGO detector for GW190521 near the event time.
}
\label{fig:denoisedH}
\end{figure}

\begin{figure}[H]
\centering
\includegraphics[clip,width=0.5\textwidth]{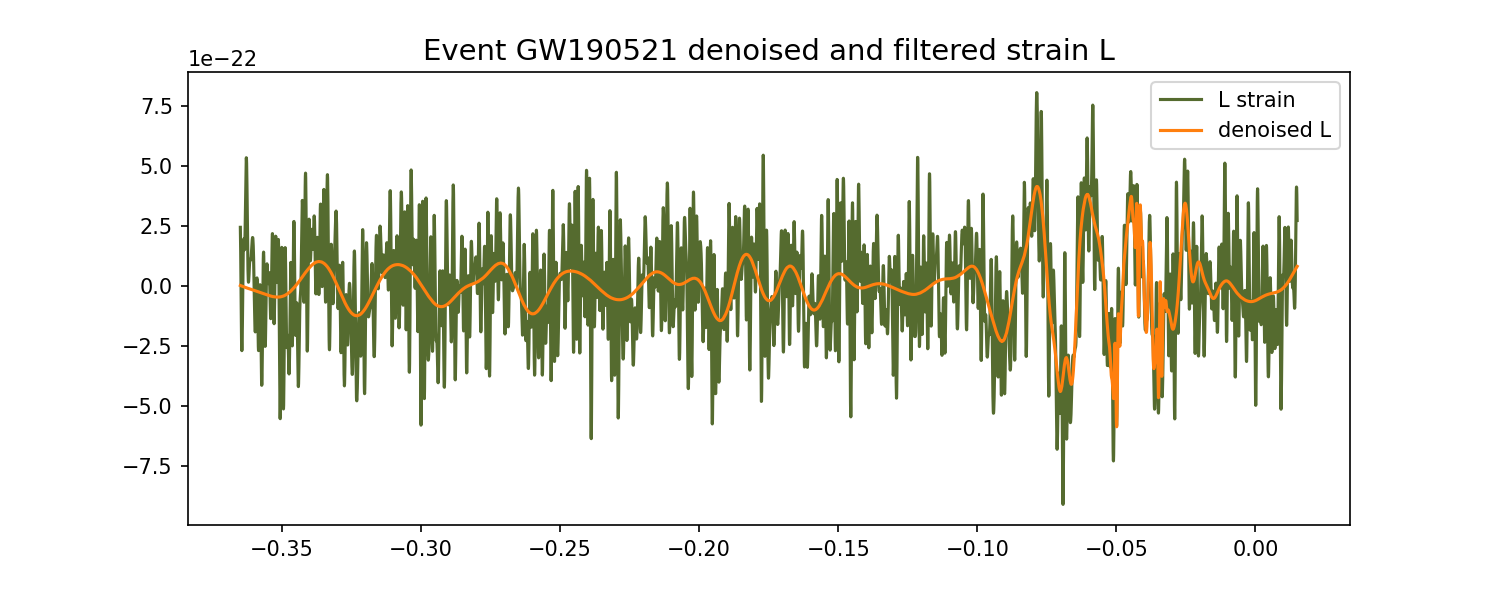}
\caption{Strain and denoised data for Livingston LIGO detector for GW190521 near the event time.
}
\label{fig:denoisedL}
\end{figure}

As described in \cite{Moreschi25a}, we apply wavelet denoising methods 
based on the general approach outlined in \cite{Mallat2009}. 
In Figs. \ref{fig:denoisedH} and \ref{fig:denoisedL}, we present 
the denoised signals for the H and L strains. 

These graphs demonstrate how the wavelet denoising process removes 
unwanted noise from the data, resulting in cleaner, more 
accurate representations of the gravitational-wave signals recorded 
during the GW190521 event.

\section{Study of the signals in the time-frequency domain}\label{sec:time_freq}

\begin{figure}[H]
\centering
\includegraphics[clip,width=0.49\textwidth]{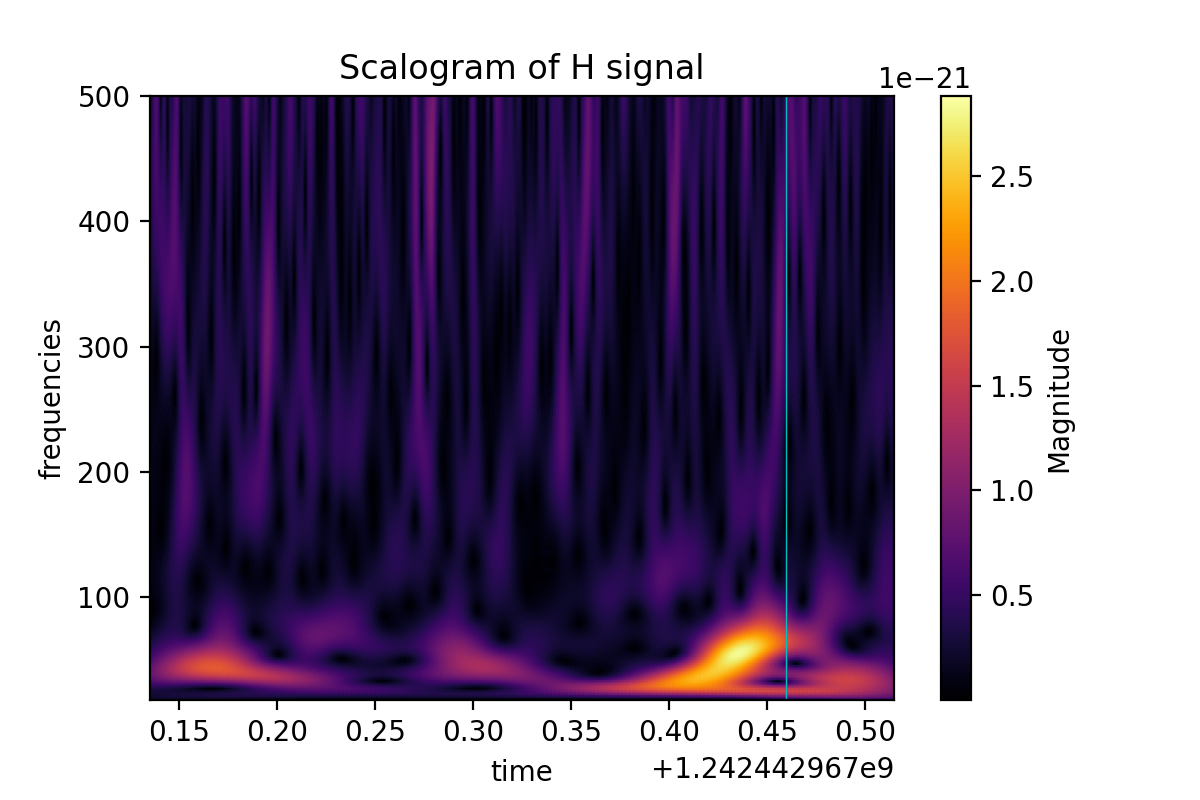}
\caption{Detail of the strain of LIGO detector H for GW190521 near the event time.
}
\label{fig:scalogramH}
\end{figure}

\begin{figure}[H]
\centering
\includegraphics[clip,width=0.49\textwidth]{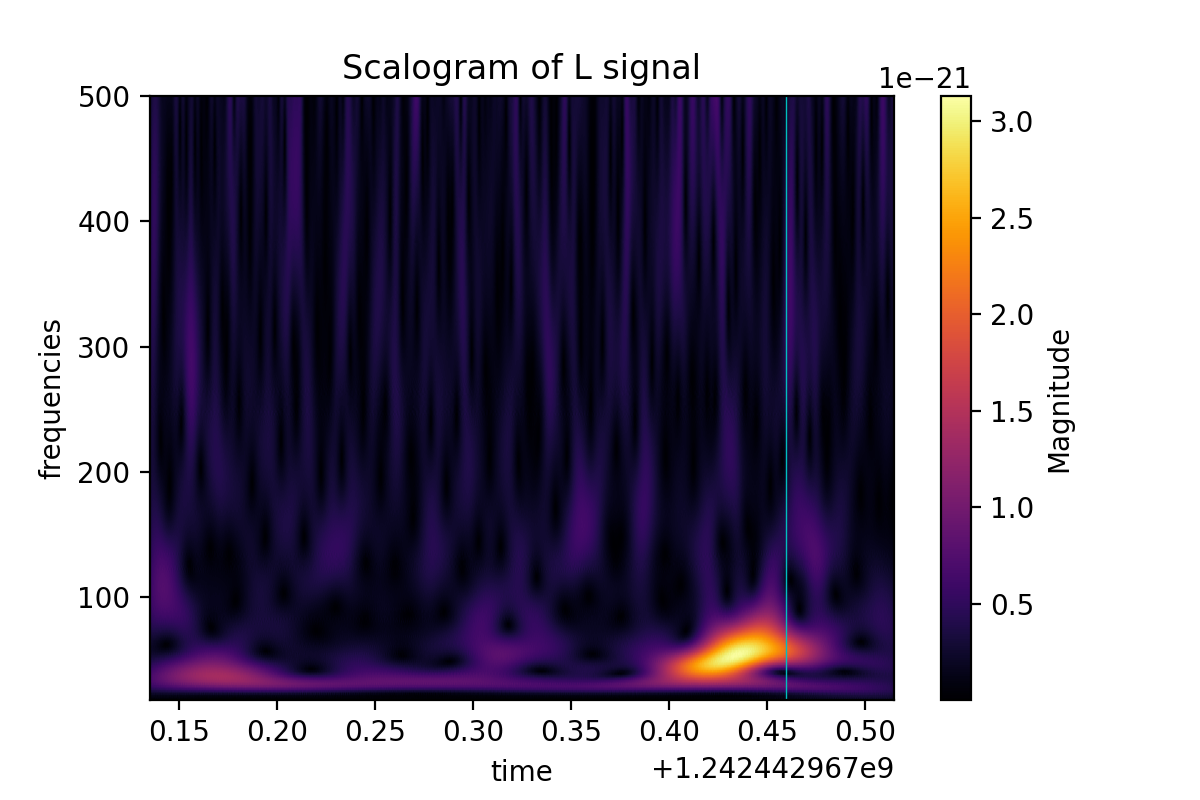}
\caption{Detail of the strain of LIGO detector L for GW190521 near the event time.
}
\label{fig:scalogramL}
\end{figure}

As part of the systematic analysis of gravitational-wave signals, 
it is common practice to examine the characteristics of the event in both 
the time and frequency domains. 
In this work we use scalograms, which allow us 
to gain insights into the signal’s structure and help estimate crucial parameters like
the final time $t_f$ and the working chirp time $t_{ch}$, 
needed for fitting in later stages.
We have found that scalograms are particularly advantageous 
for our analysis of GW events. Scalograms, which provide a 
time-frequency representation based on wavelet transforms, offer higher 
precision compared to traditional spectrograms. They allow us to 
observe the signal's frequency content and how it evolves over time 
with greater accuracy. These scalograms, for both the Hanford and Livingston detectors, 
are shown in Figs. \ref{fig:scalogramH} and \ref{fig:scalogramL}, respectively.
It can be observed a chirp like signal in both detectors involving rather low frequencies.
We have indicated with a vertical line an approximate chirp time(See discussion in \cite{Moreschi25a}).

\section{Using a universal fitting chirp form for gravitational-wave polarization modes}
\label{sec:univ_fitt_chirp}

In Figs. \ref{fig:scalogramH} and \ref{fig:scalogramL} of the scalograms of the H and L
LIGO strains for event GW190521, one can see that there is a slight increase
of frequency and strength in the signal just before the nominal event time.
For this reason, we study here the possibility of
fitting the  gravitational-wave signals with a couple of universal chirp shape
functions, that could handle a gross representation of the modes
during the inspiral phase; as described previously in \cite{Moreschi25a}.
This approach is useful for simplifying the description of the polarization modes 
during the inspiral phase and can provide a good approximation before more detailed 
models (like those from numerical relativity) become necessary in the final merger phase.

In this work we also choose the function $g(t) = 1/ \big( (t_f - t)^{p_a /4} + \epsilon_t^{p_a /4} \big)$
for fitting the amplitude time dependence of the modes,
and the function
$\Phi(t)=-2 \big(\frac{ t_f - t}{5 t_{ch}}\big)^{p_c 5/8}  + \phi_f$
for fitting the phase time dependence of the modes.
Then, we define mono-components PM as:
\begin{equation}\label{eq:sf+}
\mathsf{P}_+(t) = A_+ g(t) 
\cos( \Phi(t)  )
,
\end{equation}
and
\begin{equation}\label{eq:sfx}
\mathsf{P}_\times(t) = A_\times g(t) \sin( \Phi(t) )
,
\end{equation}
with adjustable parameters $[A_+, A_\times, \phi_f  ]$;
while the other parameters $[t_f,p_a, \epsilon_t,t_{ch},p_c]$ are fixed from the time frequency studies. 
More concretely, we define the corresponding fitting signals
\begin{equation}\label{eq:wpX}
w'_X = B_{+X} g(t) \cos( \Phi(t)) + B_{\times X} g(t) \sin( \Phi(t))
.
\end{equation}

Then, as described previously in \cite{Moreschi25a},
we can study the zeros of
\begin{equation}\label{keq:C+}
C_+(\delta,\alpha,\psi) = B_{+H}  F_{+L} - B_{+L} F_{+H}
,
\end{equation}
and
\begin{equation}\label{keq:Cx}
C_\times(\delta,\alpha,\psi) = B_{\times H}  F_{\times L} - B_{\times L} F_{\times H}
;
\end{equation}
where here the detector patern functions are thought of as functions
on the celestial sphere angles and the polarization frame angle $(\delta,\alpha,\psi)$.
In this way, for each choice of $\psi$ we study the minima of
\begin{equation}\label{eq:C+2_Cx2}
N(\delta,\alpha,\psi) =  C_+^2 + C_\times^2
,
\end{equation}
in terms of the location angles.
Also, we can study the maxima of
\begin{equation}\label{eq:invC+2_invCx2}
\mathsf{N}(\delta,\alpha,\psi) = \frac{1}{C_+^2}  + \frac{1}{C_\times^2}
;
\end{equation}
where each minimum of $C_{+,\times}$ contributes independently.

We use as initial measure the function
\begin{equation}\label{eq:M}
M_i= \frac{1}{\sqrt{N}} = \frac{1}{\sqrt{ C_+^2 + C_\times^2}}
;
\end{equation}
where the location would be indicated by the maximum values.

\begin{figure}[H]
\centering
\includegraphics[clip,width=0.5\textwidth]{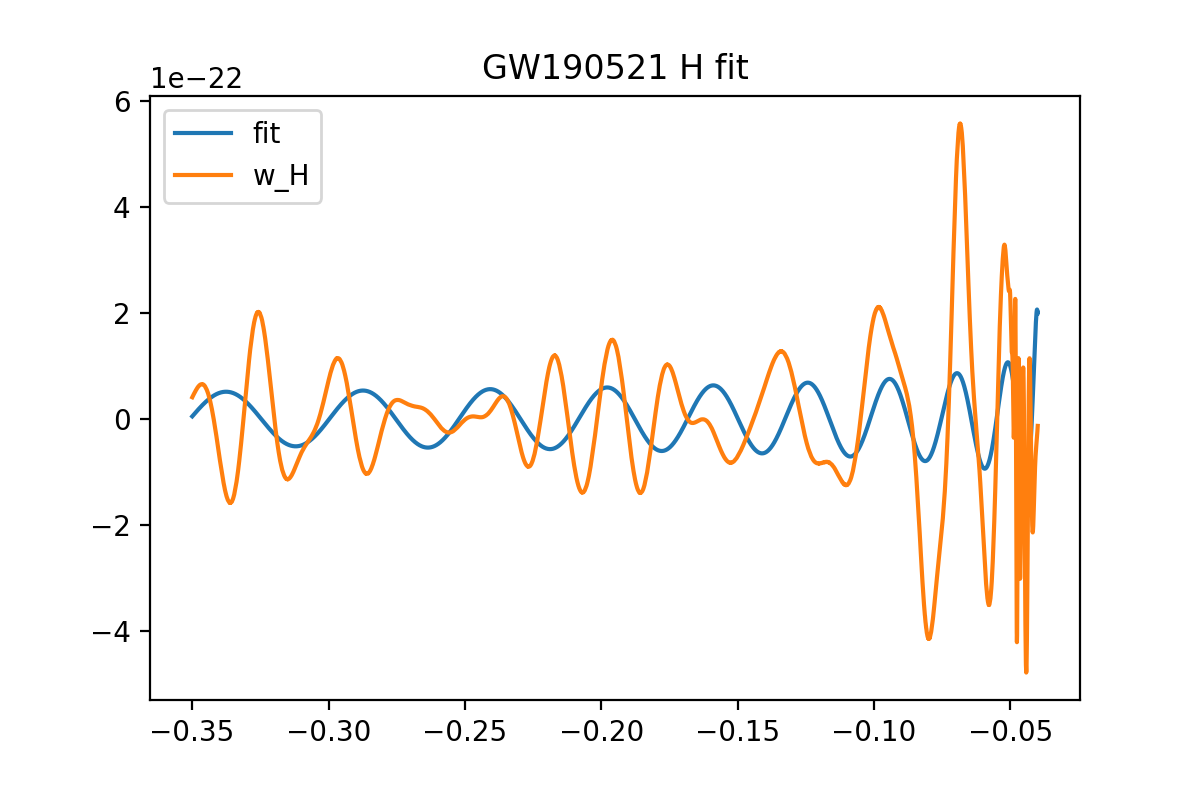}
\caption{Fitting result to the denoised signal H.
}
\label{fig:fitH}
\end{figure}

\begin{figure}[H]
\centering
\includegraphics[clip,width=0.5\textwidth]{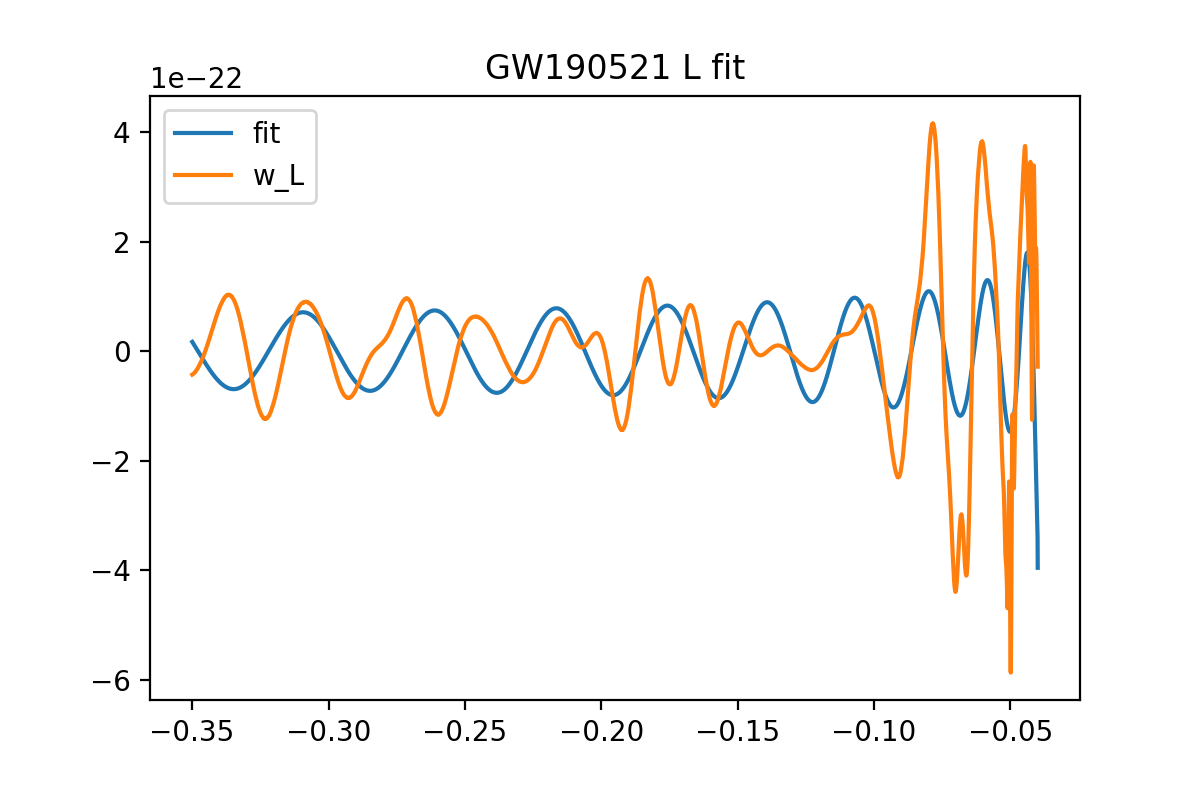}
\caption{Fitting result to the denoised signal L.
}
\label{fig:fitL}
\end{figure}

The results of fitting a universal chirp form for the polarization
of the GW to the denoised signals are shown in 
Figs. \ref{fig:fitH} and \ref{fig:fitL}.
\noindent
While using universal chirp shape functions may seem like a gross approximation 
to the signals, the surprising result is that even with these simplified representation
of the signals,
we can still obtain excellent results. This speaks to the robustness of the 
approach and highlights how effective even basic models can be in providing 
accurate localizations and polarization reconstructions for gravitational 
wave signals like GW190521.

\section{Localization of GW190521}\label{sec:loc}

\begin{figure}[H]
\centering
\includegraphics[clip,width=0.5\textwidth]{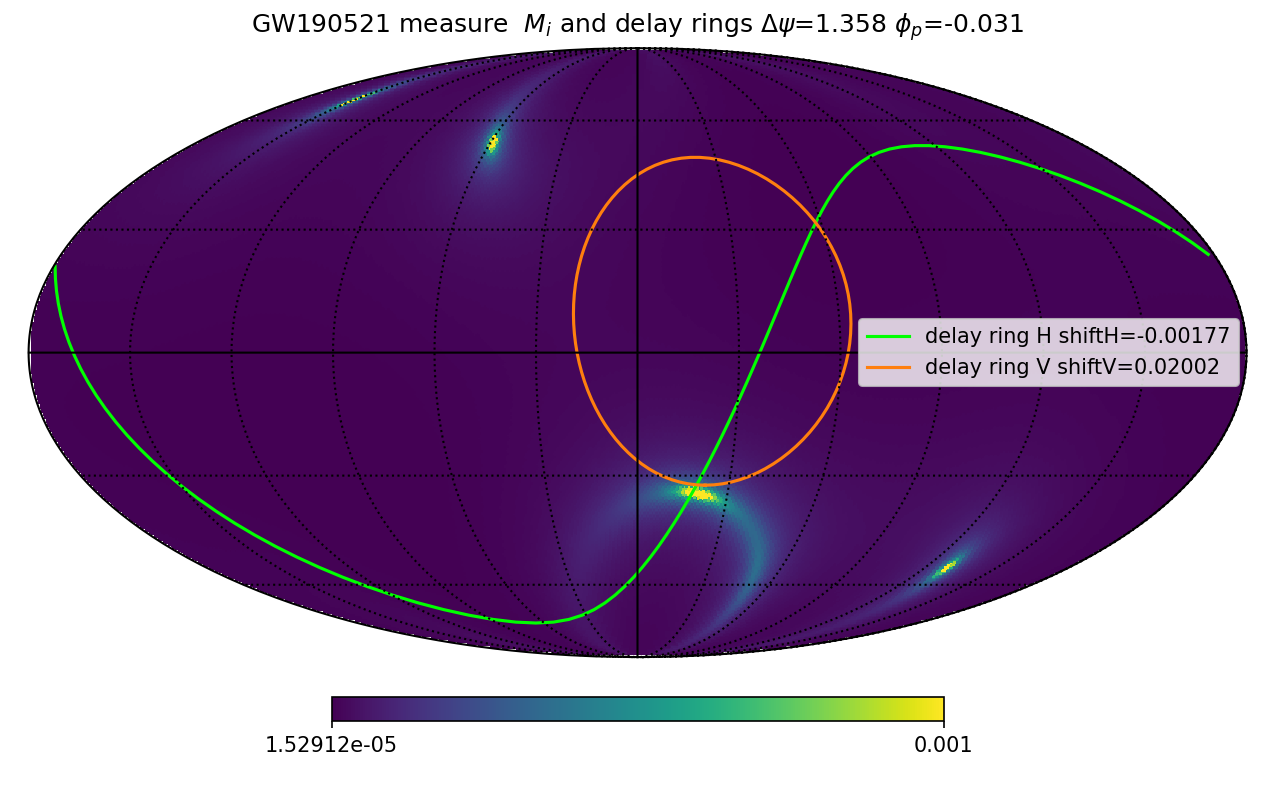}
\caption{Sky localization for the $M_i$ measure.
The delay ring for H relative to L is denoted by a lime color line.
The Virgo delay ring, also relative to L, is denoted by an orange color.
}
\label{fig:loc-M_i}
\end{figure}

The preliminary measure $M_i$ 
for the localization of event GW190521 is illustrated in Fig. \ref{fig:loc-M_i}, 
where the colored lines represent the delay rings for H relative 
to L and V relative to L. The figure 
shows a local maximum that appears near one of the intersection points of 
the delay rings, which is consistent with expectations from the triangulation method. 
This result highlights the effectiveness of the L2D+PMR procedure for localizing 
gravitational-wave sources.

\begin{figure}[H]
\centering
\includegraphics[clip,width=0.5\textwidth]{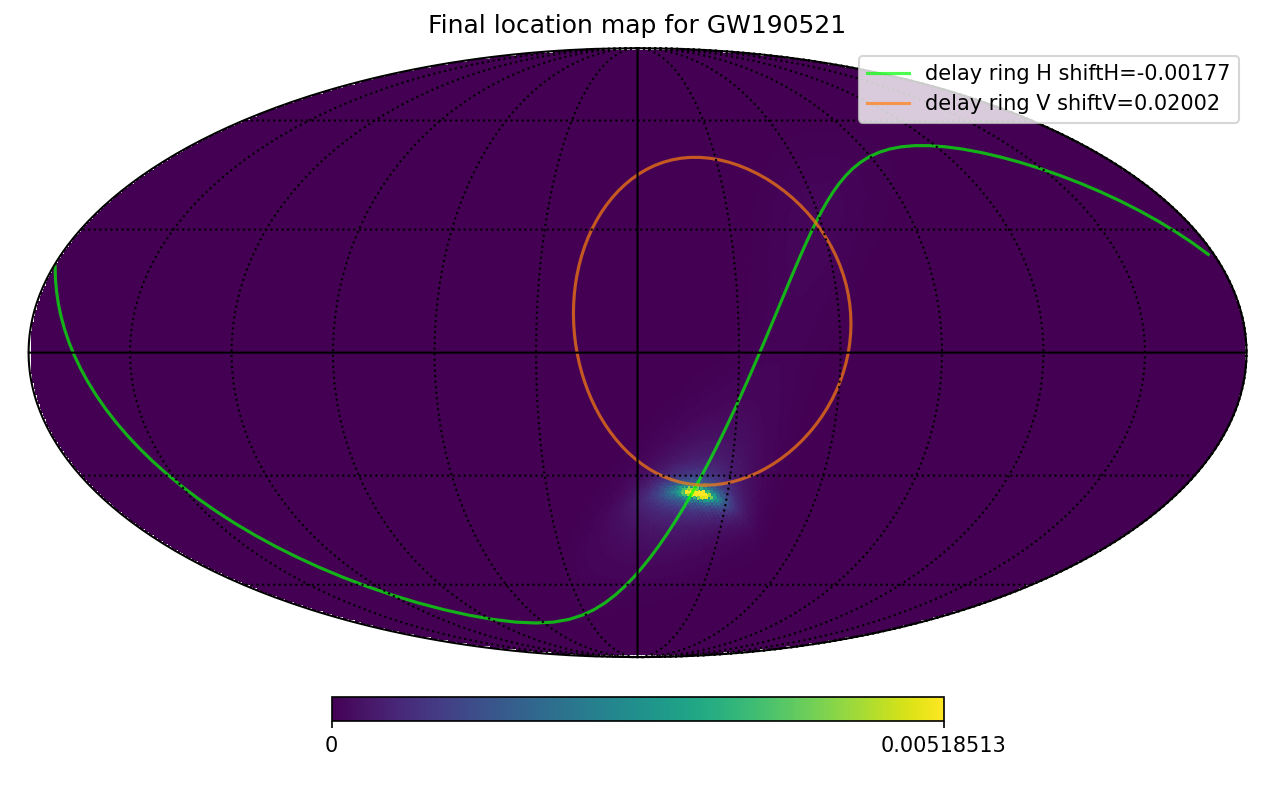}
\caption{Final sky localization for the source of GW190521,
	taking into account the measure $M_i$ and the Gaussian maps for both rings.
}
\label{fig:final_loc}
\end{figure}

\begin{figure}[H]
\centering
\includegraphics[clip,width=0.48\textwidth]{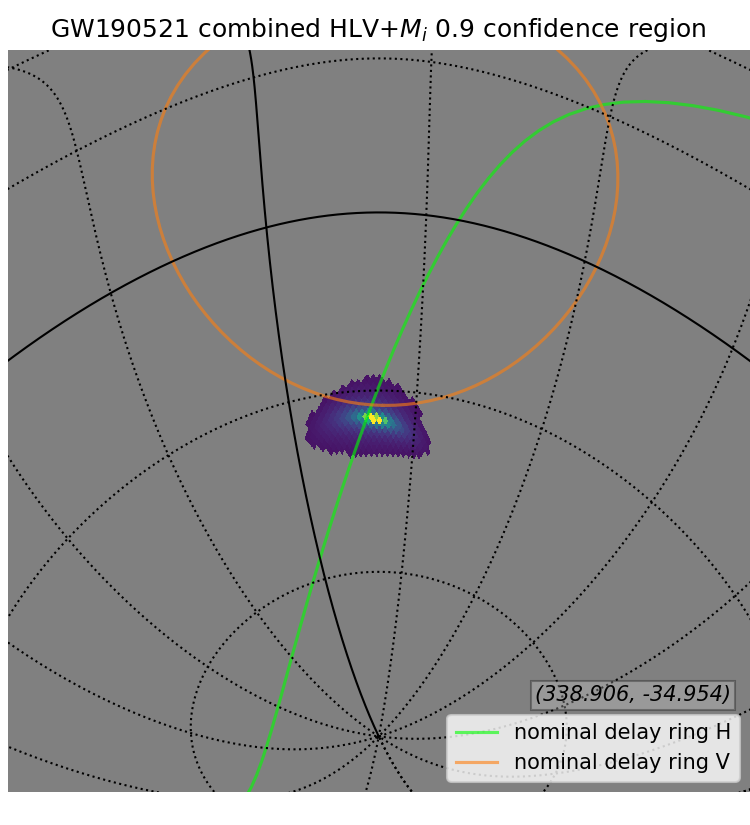}
\caption{Location at 0.9 confidence level region for the source of GW190521.
	}
\label{fig:final_loc_90pc}
\end{figure}

The final localization of the GW190521 event, as shown in Fig. \ref{fig:final_loc}, 
confirms that the L2D+PMR procedure is indeed consistent with expectations. 
The location is very close to one of the intersections of the delay rings from the 
H-L and V-L delay rings. This alignment between the calculated localization and 
the expected region near the delay ring intersections serves as a validation  
of the procedure, highlighting its potential as a reliable 
tool for gravitational-wave source localization.
Figure \ref{fig:final_loc_90pc} presents the corresponding 0.9 confidence region, 
constructed following the methodology described in \cite{Moreschi25a}.


For comparison we also show here the localization published by the LIGO team for this event
\begin{figure}[H]
\centering
\includegraphics[clip,width=0.49\textwidth]{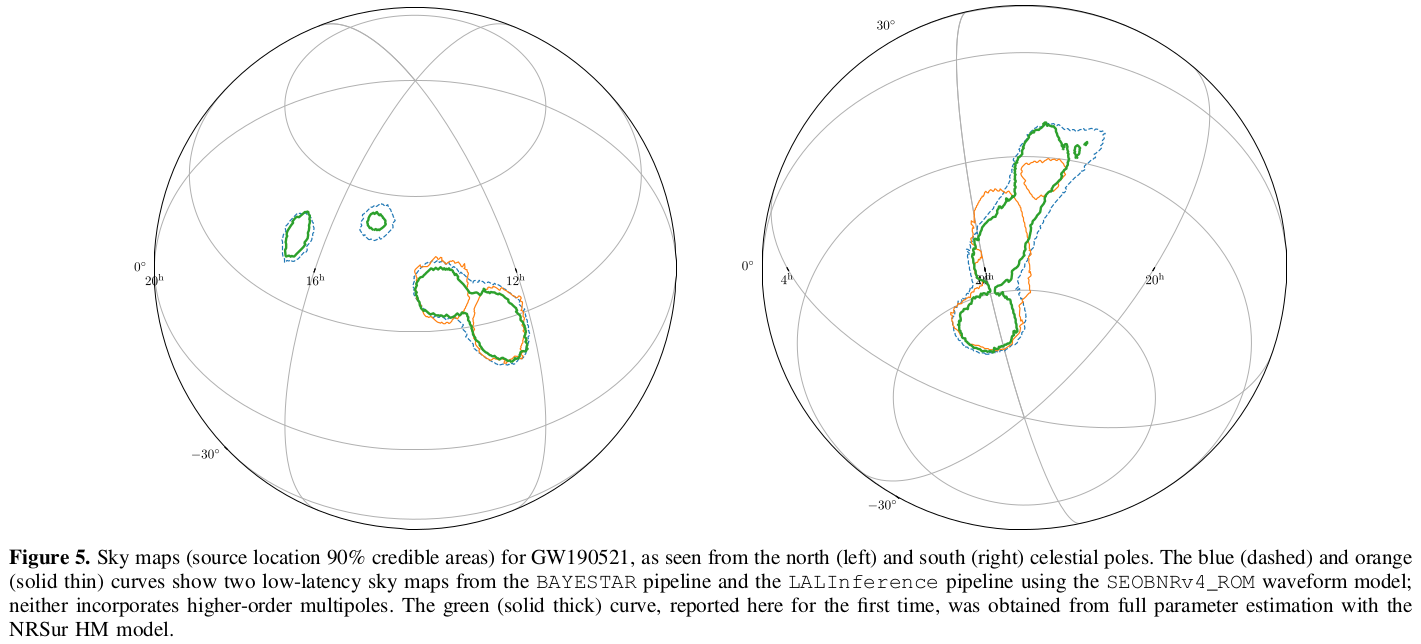}
\caption{This is the reproduction of figure 5 from reference \cite{LIGOScientific:2020ufj}
where the LIGO team shows the sky localization at 90\% credible areas for GW190521 using the Bayestar and LALInference pipelines,
and also the results of a full parameter estimation with the NRSur HM model.
}
\label{fig:loc-ligo-bayestar+LALI}
\end{figure}

\begin{figure}[H]
\centering
\includegraphics[clip,width=0.49\textwidth]{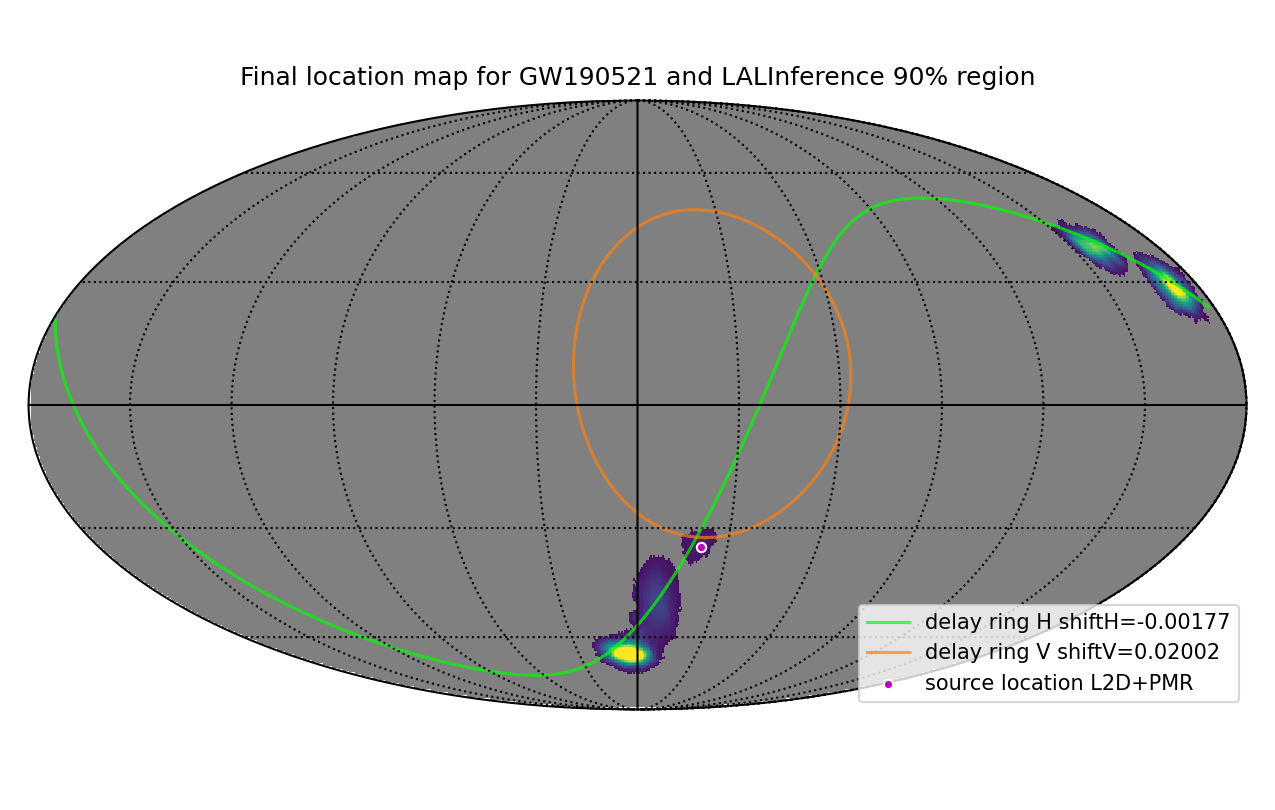}
\caption{Sky localization from LIGO team with LALInference method, our localization 
and the reference delay rings.
The colored regions with grades of blue, green and yellow, corresponds to the enclosed
area by orange curves from the previous LIGO graphs, showing the localization sky map
using the LALInference pipeline; where yellow indicates more intensity.
The magenta dot indicates our best localization for the source.
The H-L and V-L delay rings are overlaid to show the results from the time-delay triangulation
}
\label{fig:loc-ligo-LALI+nuestro+rings}
\end{figure}

In \cite{LIGOScientific:2020ufj} the authors reported that Bayestar\citep{Singer:2015ema} pipeline provides a 90\%
credible area of 1163deg$^2$; while the LALInference analysis\citep{Veitch:2014wba} provides a 90\%
credible localization within 765deg$^2$.
The 0.9 confidence region in our graph of Fig. \ref{fig:final_loc_90pc}
covers an area of 250deg$^2$.
This means that we provide with a more precise localization procedure;
but it remains the question of the accuracy of the methods.
To investigate this,
we show in Fig. \ref{fig:loc-ligo-LALI+nuestro+rings} the LALInference 90\% region, our best
localization point and the two reference delay rings H-L and V-L.
It can be seen in Fig. \ref{fig:loc-ligo-LALI+nuestro+rings} that the sky map source location
at 90\% credible area for GW190521 using the LALInference pipeline appears separated
in four regions. The magenta dot is centered at the best location, as suggested
by the L2D+PMR procedure.  
It is observed that our best location is within one of the LALInference regions;
and that both are very closed to one of the crossings of the two delay rings.
Recalling that the factually accurate location should be at one of the crossings
of the two delay rings, we deduce that all the methods considered show
signals close to the southern crossing, and none close to the northern one.
Therefore we conclude that LALInference pipeline and the L2D+PMR procedure
have similar very good accuracy for this event.
Since the other two methods behave similarly as the LALInference pipeline;
we also conclude that they also show very good accuracy for this event.

\section{Reconstruction of the spin-2 polarization modes of GW190521}\label{sec:PM}

\subsection{In the working polarization frame}

\begin{figure}[H]
\centering
\includegraphics[clip,width=0.49\textwidth]{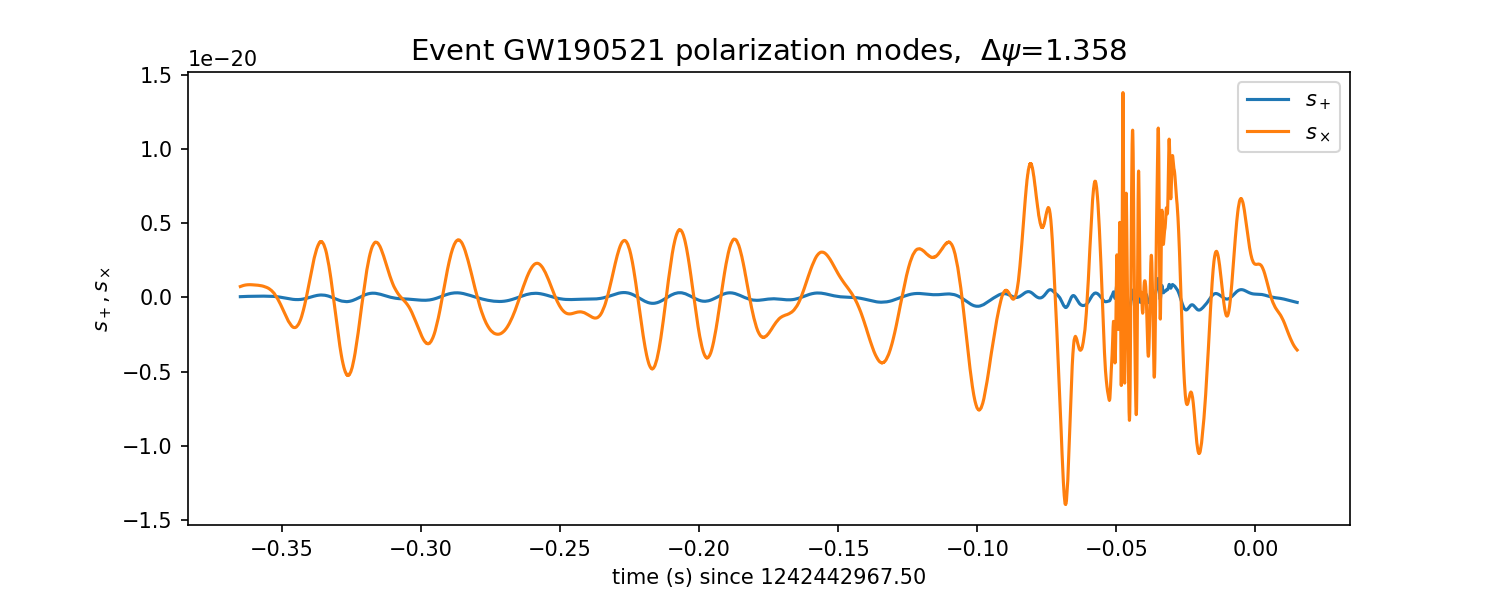}
\caption{Polarization modes + and $\times$ close to the reference time of the GW190521 event.}
\label{fig:s+_sx}
\end{figure}

With the localization of the source in the celestial sphere now determined, 
the next crucial step is the reconstruction of the polarization modes 
of the gravitational-wave signal. As shown in Fig. \ref{fig:s+_sx}, 
the + and $\times$ PMs are reconstructed from the 
gravitational-wave signals detected by the LIGO detectors. 
It is remarkable in Fig. \ref{fig:s+_sx} that at the frame $\Delta \psi = 1.358$
there is almost no contribution for the plus polarization mode.
This means that the sky location is such that both LIGO detectors have
recorded almost the same component of the PM of the GW;
more precisely -H strain has recorded almost the same polarization mode component 
as the L strain.

\begin{figure}[H]
\centering
\includegraphics[clip,width=0.49\textwidth]{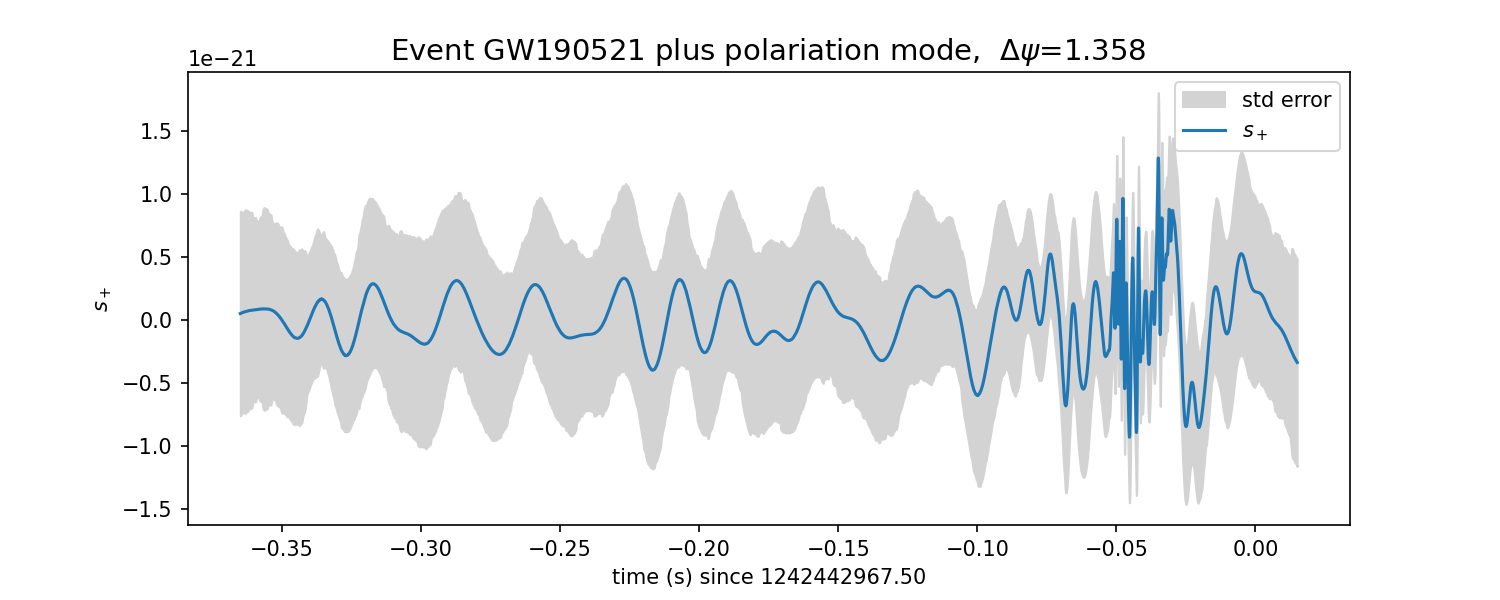}
\includegraphics[clip,width=0.49\textwidth]{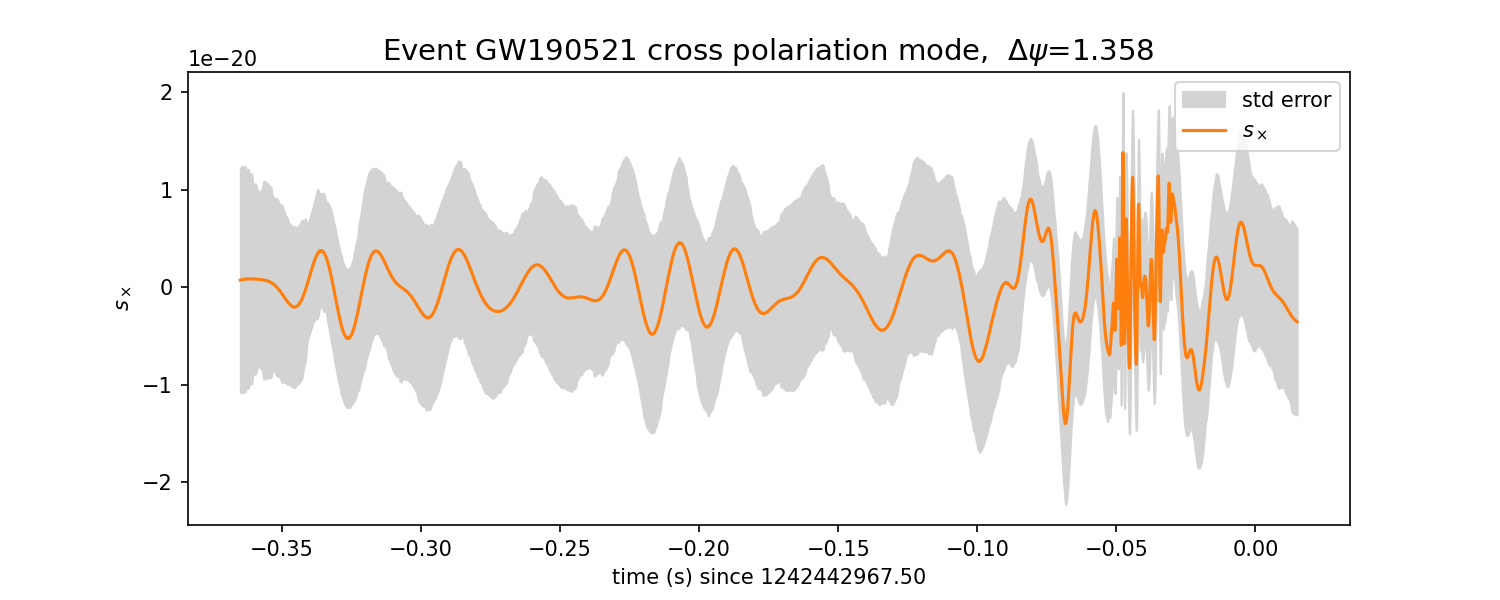}
\caption{Polarization modes of the event GW190521 for $\Delta\psi=1.358$ with estimated error bands,
	in the region close to the nominal event time.
}
\label{fig:s+_sx_errores}
\end{figure}

In Fig. \ref{fig:s+_sx_errores}
we show the graphs of the polarization modes with their respective upper bound
estimated error bands; where it can be noticed that
the small $s_+$ is very similar in shape (not in magnitude) to $s_\times$;
which is consistent with the previous conjecture that both observatories
have detected essentially the same polarization component of the GW.

The nearly identical nature of the modes suggests that the sky location of the source 
was such that the polarization modes were primarily oriented in a direction 
that allowed both detectors to register almost the same PM of 
the gravitational wave.
The observed error bands also help to assess the precision of the polarization 
reconstruction.
The consistency between the modes, even with the error bands considered, 
suggests that the L2D+PMR procedure provides an accurate and reliable 
reconstruction of the GW’s polarization properties.
We use the same estimation of the errors for the spin-2 polarization modes
as described in \cite{Moreschi25a}.

\subsection{Spin-2 polarization modes of GW190521 for different polarization angles}

\begin{figure}[H]
	\centering
	\includegraphics[clip,width=0.49\textwidth]{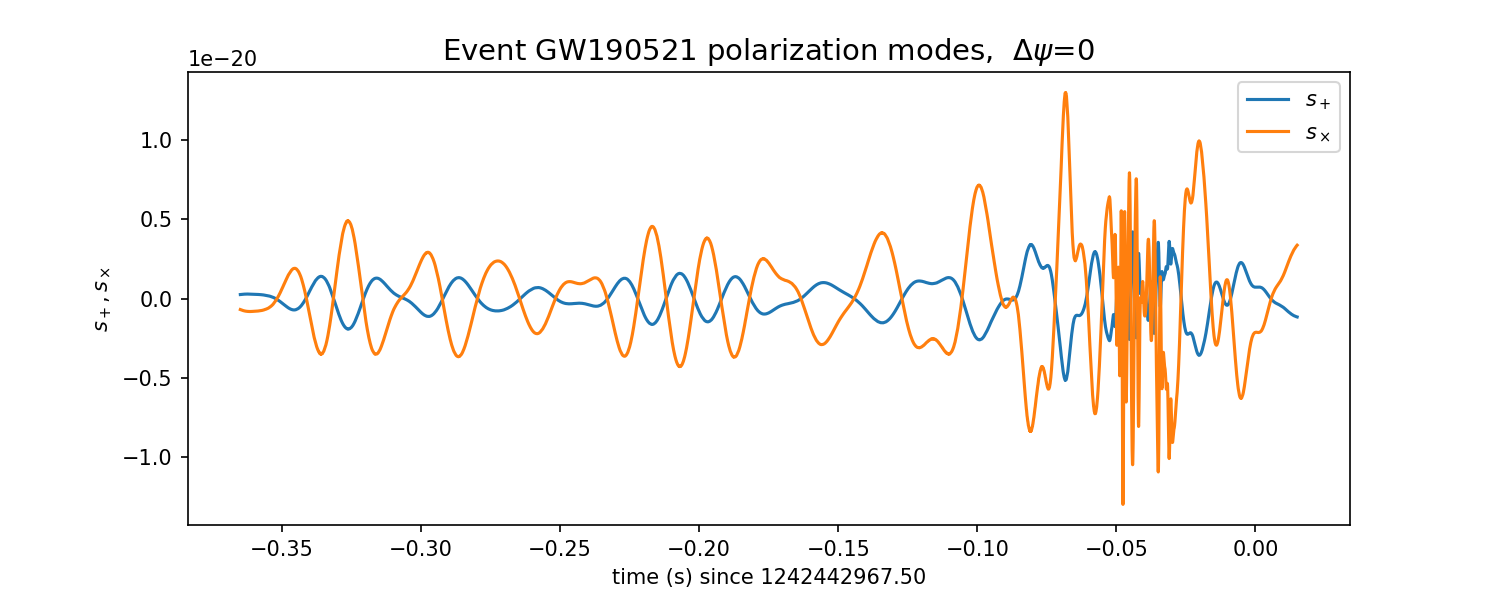}
	\caption{Polarization modes of the event GW190521 for $\Delta\psi=0$.
	}
	\label{fig:s+_sx_4-0}
\end{figure}

\begin{figure}[H]
	\centering
	\includegraphics[clip,width=0.49\textwidth]{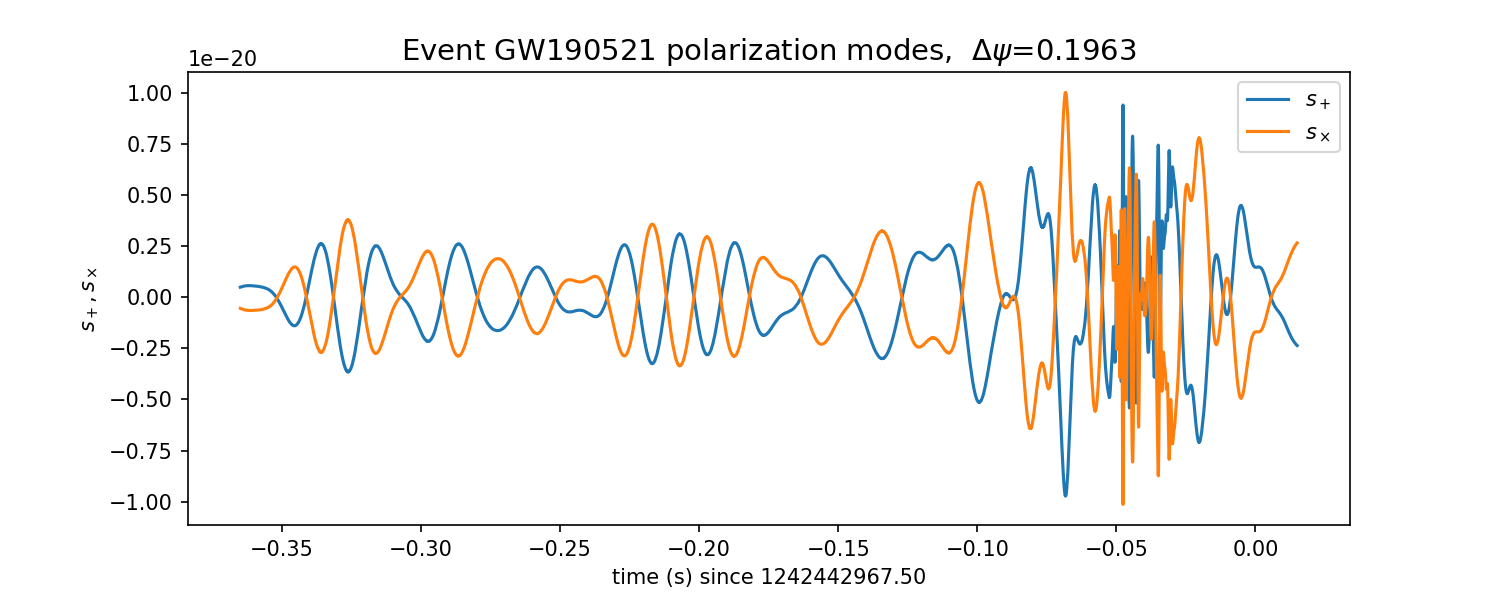}
	\caption{Polarization modes of GW190521.
	}
	\label{fig:s+_sx_1}
\end{figure}

\begin{figure}[H]
	\centering
	\includegraphics[clip,width=0.49\textwidth]{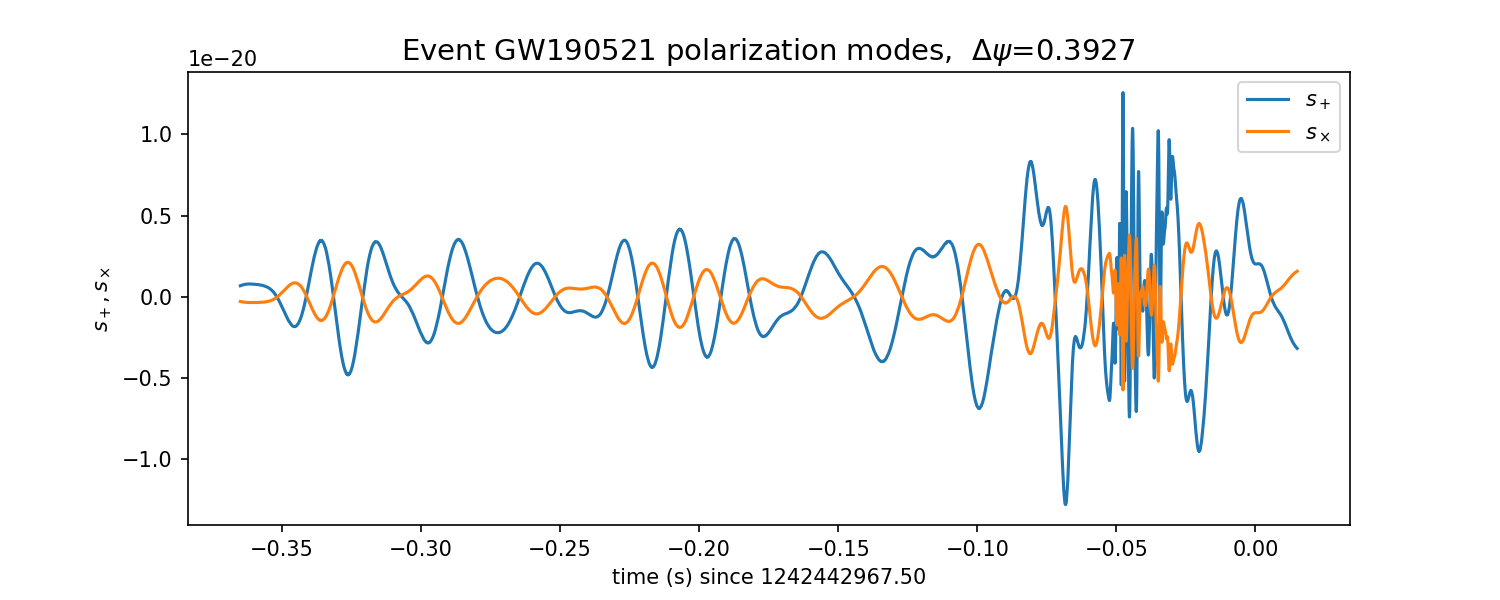}
	\caption{Polarization modes of GW190521.
	}
	\label{fig:s+_sx_2}
\end{figure}

\begin{figure}[H]
	\centering
	\includegraphics[clip,width=0.49\textwidth]{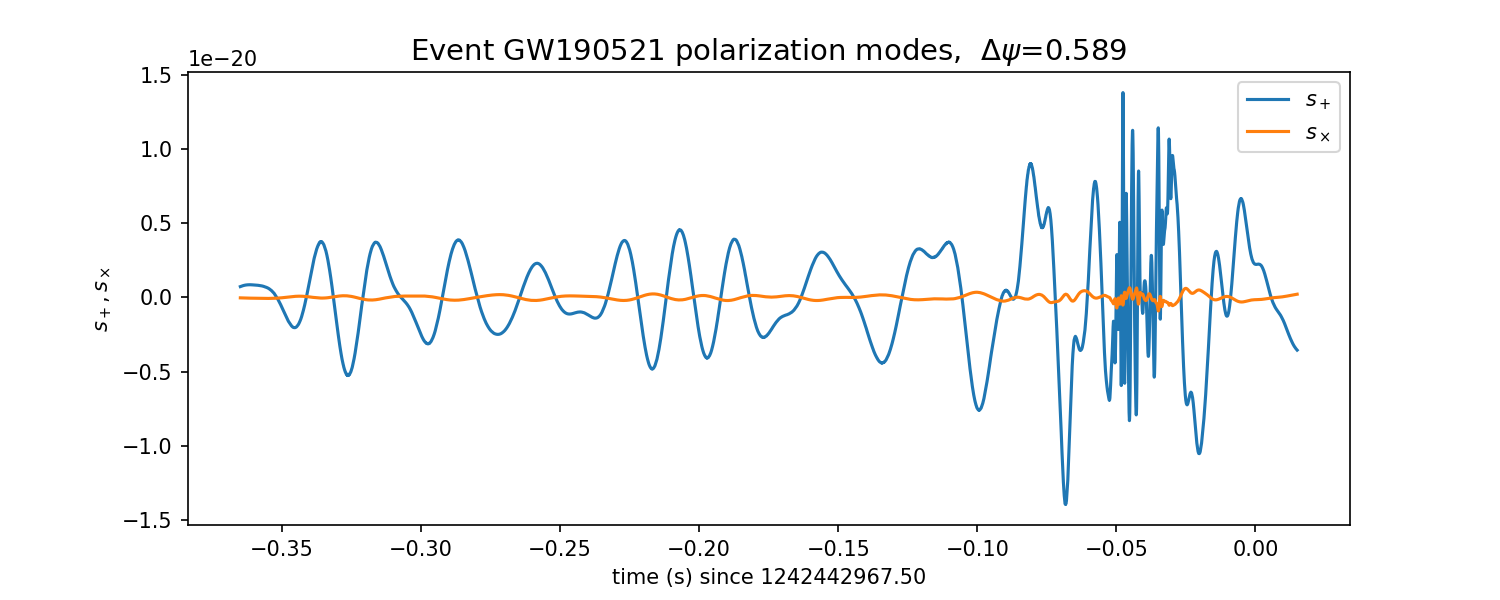}
	\caption{Polarization modes of GW190521.
	}
	\label{fig:s+_sx_3}
\end{figure}

\begin{figure}[H]
	\centering
	\includegraphics[clip,width=0.49\textwidth]{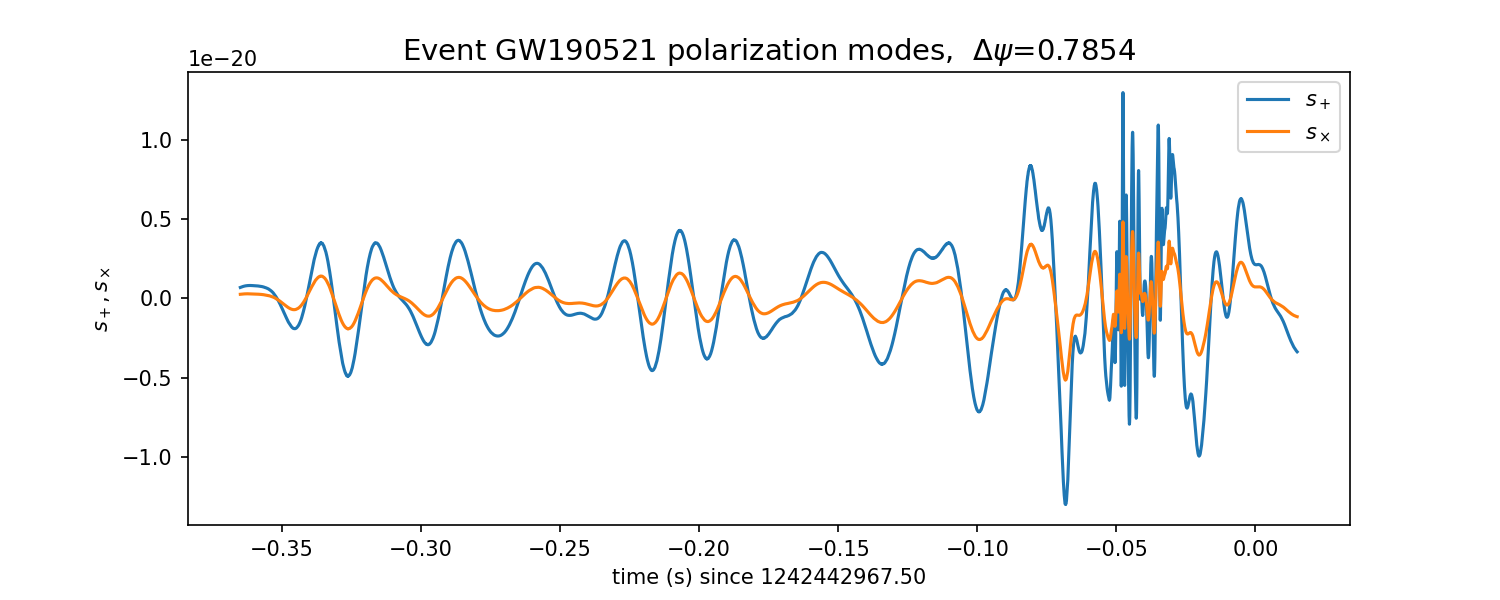}
	\caption{Polarization modes of GW190521.
	}
	\label{fig:s+_sx_4}
\end{figure}

In Figs. \ref{fig:s+_sx_4-0}-\ref{fig:s+_sx_4}
we show the graphs for the polarization modes in the frames described by the choices
$\Delta\psi=0$, $\Delta\psi=\frac{\pi}{16}$, $\Delta\psi=\frac{2\pi}{16}$, $\Delta\psi=\frac{3\pi}{16}$ and $\Delta\psi=\frac{\pi}{4}$. 
It can be seen that for $\Delta\psi=\frac{\pi}{4}$ the PM return to
the original values according to the transformation properties of the modes.

It can be observed that the polarization modes in each frame do not seem to be
independent. In fact, calculating the correlation $\rho_{s_+,s_\times}$
in each case, one finds:
for $\Delta\psi=0$, $\rho_{s_+,s_\times}=-0.9945$;
for $\Delta\psi=\frac{\pi}{16}$,  $\rho_{s_+,s_\times}=-0.9976$;
for $\Delta\psi=\frac{2\pi}{16}$, $\rho_{s_+,s_\times}=-0.9958$;
for $\Delta\psi=\frac{3\pi}{16}$, $\rho_{s_+,s_\times}= -0.6901$;
and for  $\Delta\psi=\frac{\pi}{4}$, $\rho_{s_+,s_\times}=0.9945$.
It is concluded then that the PM in all frames are correlated
with the exception of the case $\Delta\psi=\frac{3\pi}{16}$, but which corresponds
to Fig. \ref{fig:s+_sx_3} where one can see that one of the components is
very small, which in turn produces big numerical errors.

The LIGO/Virgo Collaborations acknowledge\citep{KAGRA:2023pio} hardware injections in the
Advance LIGO and Advance Virgo detector during the run O3; and
checking the segment lists showing times when injections were not present,
it is noted that the event time of GW190521 is not in the
NO\_CW\_HW segment lists, which show times when Continuous Wave (CW) injections were not present.
The detailed account of injections during the O3 run is described
at \url{http://gwosc.org/O3/o3_inj}.
Since the CW injection corresponds to a simulation of a gravitational-wave signal 
of the form expected from an isolated neutron star;
the physical parameters of the injected signal are very different
to those measured in the GW190521 event.
That is, there is no danger in misinterpretation of these signals
to be confused with the result of a one component injection.

All this reinforces the conjecture that both LIGO observatories have detected
in this case essentially the same polarization component.
What is somehow striking is that in spite of this fact, the procedure 
still succeeded in accurately localizing the source.

\section{Reconstruction of the two LIGO signals in terms of just the spin-2 polarization modes}\label{sec:rec-signalfromPM}

By reconstructing the spin-2 polarization mode contributions and subtracting them from 
the original strains, we can test whether additional signal components remain, potentially 
indicating contributions from other spin sources (e.g., spin-0 or spin-1 modes).

The OM measure $\Lambda$\citep{Moreschi:2024njx}
provides a quantitative way to compare the original filtered strains with the residual 
strains after subtraction. If no significant remnant signal is found, this would support 
the assumption that the detected gravitational wave was purely spin-2 in nature. 
Conversely, if residual signal components persist after subtraction, it could indicate 
possible alternative contributions, let us say of spin 0 or 1, or systematic effects 
not accounted for in the standard spin-2 framework.

\begin{figure}[H]
\centering
\includegraphics[clip,width=0.49\textwidth]{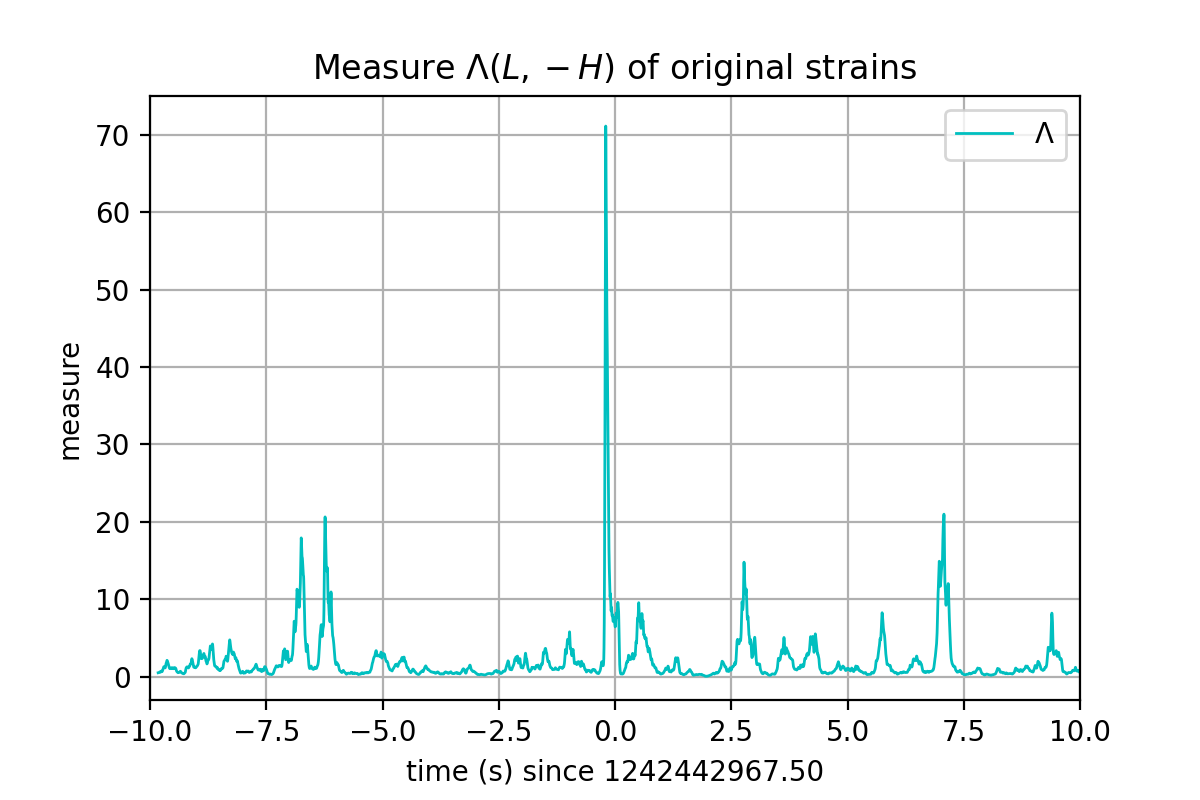}
\includegraphics[clip,width=0.49\textwidth]{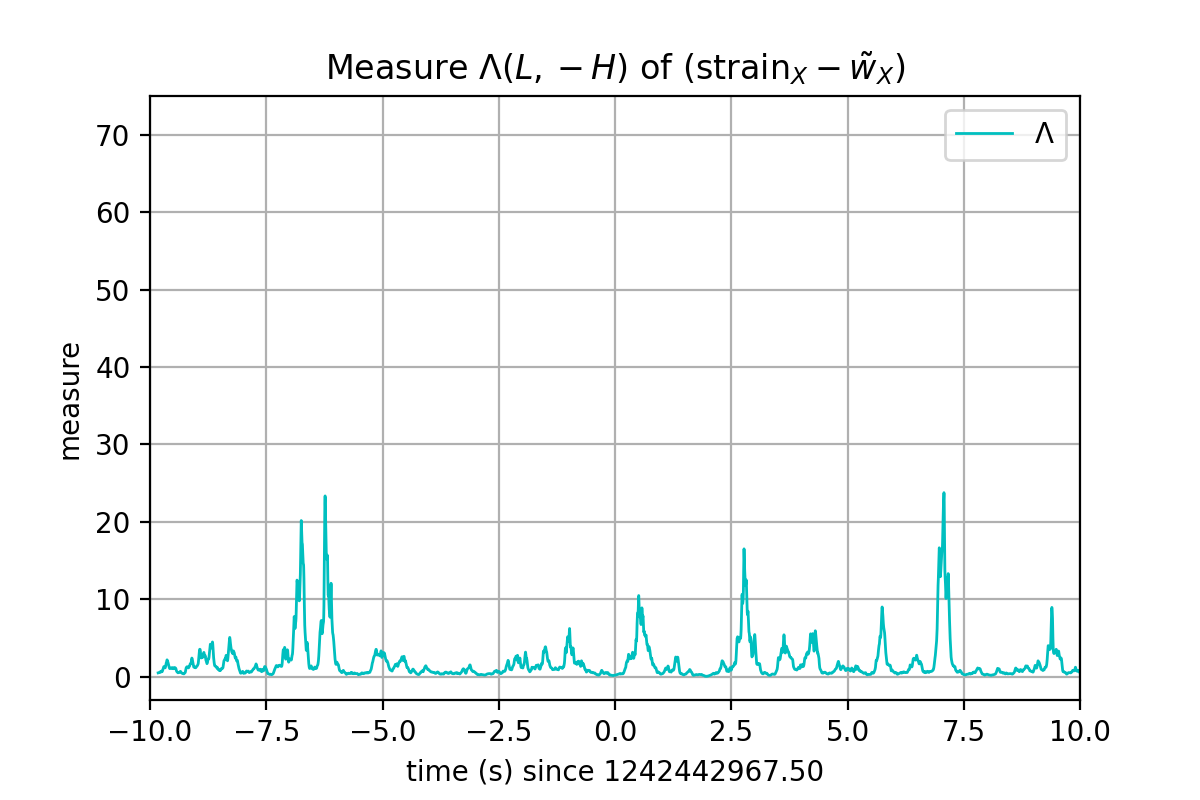}
\caption{On the top graph, the values of the measure $\Lambda$ close to the reference event time
	for the original filtered strains of GW190521,
	and on the bottom for the strains after the subtraction of the reconstructed
	signals from the spin-2 PM. The residual is consistent with the noise of
	a wide temporal window.
}
\label{fig:strain-wtilde}
\end{figure}

We present the graphs of the OM measure $\Lambda$
for the original filtered strains and for the strain after the subtraction of the reconstructed
signals from the spin-2 polarization modes in Fig. \ref{fig:strain-wtilde}.
This result strongly supports the standard prediction of general relativity, 
which states that GWs should only contain spin-2 polarization modes. 
Since the OM measure behaves as ambient noise after subtracting the reconstructed 
spin-2 signals, it indicates that no significant residual signal remains that could 
suggest the presence of spin-1 (vector) or spin-0 (scalar) polarization modes.

This finding aligns with previous gravitational-wave detections, where no deviations 
from general relativity have been observed in terms of additional polarization modes. 
The absence of spin-0 or spin-1 components in GW190521 reinforces the robustness of 
the standard model of GWs as coming from a concentrated 
region (See discussions in \cite{Moreschi25a}).

\section{Final comments}\label{sec:final}

The GW190521 event has various characteristic aspects that has
provoked the study of several subjects.
In particular,
the search for possible candidate electromagnetic counterparts for the
detected GW of GW190521 has been studied in several publications;
although in some of them they refer to the name S190521g; which is the label
used at the GraceDB\href{https://gracedb.ligo.org/superevents/S190521g/}{GraceDB.GW190521}.
Here we mention a non-exhaustive list of references related to this topic.
The search using the Fermi-LAT space telescope data, as reported in \cite{Podlesnyi:2020dbm}
found no significant signal linked to GW190521.
In reference \cite{Graham:2020gwr} they studied the alert ZTF19abanrhr from the
Zwicky Transient Facility, which was announced 34 day after the GW event,
and associated with AGN J124942.3 + 344929 at $z=0.438$.
The authors showed a graph where the location of AGN J124942.3 + 344929
appears within one of the 90\% LIGO regions; but there is no mention
in the graph of the essential delay rings, and it can be seen that
the location of this AGN is far from both crossing points of the 
delay rings we have discussed above.
Furthermore, in references \cite{Ashton:2020kyr,Palmese:2021wcv}
they consider that there is insufficient evidence to warrant confidently 
association of GW190521 with ZTF19abanrhr.
In reference \cite{Adriani:2022zli} the authors reported on the results
of a search for X-ray/gamm-ray counterparts to gravitational-wave events announced during
the O3 LIGO/Virgo observing run, using the CALorimetric Electron Telescope (CALET),
and concluded that no events have been detected that pass all acceptance criteria.

In this article we have presented the result on the localization of the source
of the GW190521 event using the L2D+PMR procedure with the H and L strains.
We have not used the V strain in the localization and reconstruction steps due to its noisy character;
but we have used the relative time information for refining the location.
In this process, we had to measure for the first time
the relative time delays H-L and V-L for this case; which allowed us
to calculate the two corresponding delay rings.
As expected, our location turns out to be very close to one of the two crossing
points of both delay rings. This indicates the consistency of the L2D+PMR methods
for the localization of gravitation-wave sources.
Our location for GW190521 is within one of the 90\% regions of the
Bayestar and LALInference LIGO pipelines, confirming the consistency of our procedure
with these two methods for this event.

The analysis of the spin-2 polarization modes reveals an interesting feature: 
both LIGO detectors recorded nearly the same polarization component of the 
GW. 
We have reconstructed the spin-2 polarization modes in multiple polarization frames, 
confirming the robustness of the results.
It should be remarked also that the sky location is responsible for
the fact that both LIGO observatories have observed this similar polarization
component; this will be the subject of further studies in future articles.

Our study further strengthens the conclusion that the gravitational-wave signal 
from GW190521 can be entirely explained by spin-2 polarization modes, with 
no significant contribution from alternative modes (such as spin-0 or spin-1). 
This is an important validation of General Relativity's prediction that 
GWs should be purely tensorial (spin-2).
The confirmation comes from the fact that
the reconstructed waveforms, when subtracted from the original data, 
leave only ambient noise, suggesting that additional polarizations are not needed.
This result aligns with previous findings\citep{Moreschi25a}.

In this article, we have deliberately chosen not to focus on the detailed astrophysical 
interpretation of our results. Instead, our primary objective has been to apply
and examine a novel procedure that enables the localization of a gravitational-wave source 
using data from only two detectors. Additionally, this method allows for the direct 
reconstruction of polarization modes from the observed data.
By concentrating on the methodological advancements, we aim to establish a robust and 
generalizable framework that can be applied to future GW detections. 
The astrophysical implications of these findings, while undoubtedly significant, 
are beyond the scope of this work and are best addressed in future studies that 
build upon the techniques introduced here.

We intend to continue the development of the L2D+PMR procedure by applying it
to other GW events with high amplitude signals.






\subsection*{Acknowledgments}

This work is possible thanks to the open data policy of the 
LIGO Scientific Collaboration, the Virgo Collaboration and Kagra Collaboration;
who are giving freely access to data through the Gravitational Wave Open Science
Center at \href{https://gwosc.org}{https://gwosc.org};
which is described in \cite{LIGOScientific:2019lzm} and \cite{KAGRA:2023pio}.

We have used python tools included in the project PyWavelets\citep{Lee:2019}

We acknowledge support from  SeCyT-UNC and CONICET.

We are very grateful to Emanuel Gallo for a careful reading
of the manuscript and for indicating several improvements.


%




\end{document}